\RequirePackage{fix-cm}
\documentclass{svjour3}                     
\smartqed  
\usepackage{graphicx}
\usepackage{amsmath}
\usepackage{amsfonts}
\usepackage{url}
\usepackage{hyperref}
\usepackage{physics}
\usepackage{color}
\usepackage{subcaption}
\usepackage{multirow}
\usepackage[framemethod=tikz]{mdframed}
\usepackage{xcolor}
\usepackage{mathrsfs} 
\usepackage{longtable}
\usepackage{afterpage}

\renewcommand{\order}[1]{O\left( #1 \right)}

\newcommand{\dt}{\partial_t}

\newcommand{\p}{\partial}
\newcommand{\mrs}{\mathrm{s}}
\newcommand{\mre}{\mathrm{e}}

\journalname{Journal of Engineering Mathematics}
\begin{document}

\title{Derivation of an Effective Thermal Electrochemical Model for Porous Electrode Batteries using Asymptotic Homogenisation\thanks{The authors acknowledge support from the EPSRC Prosperity Partnership (EP/R004927/1) and The Faraday Institution (EP/S003053/1 grant number FIRG003).}
}

\titlerunning{Homogenisation of Thermal Electrochemical Model for Batteries}        

\author{M. J. Hunt         	\and
        F. Brosa Planella	\and
        F. Theil			\and
        W. D. Widanage
}


\institute{M. J. Hunt \and F. Brosa Planella \and W. D. Widanage \at
              WMG, University of Warwick, Coventry CV4 7AL, UK \\
              \email{Ferran.Brosa-Planella@warwick.ac.uk}           
           \and
           M. J. Hunt \and F. Theil \at
				Mathematics Institute, University of Warwick, Coventry CV4 7AL, UK \\
			\and
		F. Brosa Planella \and W. D. Widanage \at
			The Faraday Institution, Quad One, Becquerel Avenue, Harwell Campus, Didcot, OX11 0RA, UK
}

\date{Received: date / Accepted: date}

\maketitle

\begin{abstract}
Thermal electrochemical models for porous electrode batteries (such as lithium ion batteries) are widely used. Due to the multiple scales involved, solving the model accounting for the porous microstructure is computationally expensive, therefore effective models at the macroscale are preferable. However, these effective models are usually postulated ad hoc rather than systematically upscaled from the microscale equations. We present an effective thermal electrochemical model obtained using asymptotic homogenisation, which includes the electrochemical model at the cell level coupled with a thermal model that can be defined either at the cell or the battery level. The main aspects of the model are the consideration of thermal effects, the diffusion effects in the electrode particles, and the anisotropy of the material based on the microstructure, all of them incorporated in a systematic manner. We also compare the homogenised model with the standard electrochemical Doyle, Fuller \& Newman model.
\end{abstract}

\section{Introduction}
The role of rechargeable batteries has become more and more important in recent years due to the increase in the use of electronic devices and electric vehicles. In particular, most of these applications rely on lithium ion batteries, a type of porous electrode battery, and therefore research on this type of battery has received a lot of attention in the past few years. The first electrochemical model for porous electrode batteries was put forward in the seminal paper of Doyle, Fuller and Newman \cite{Fuller1994}. In that article, an effective model was presented which described the mass and charge transport in both the electrode and the electrolyte. The porous structure of the electrode was modelled by assuming that at each point of the electrodes there is a representative particle in which lithium intercalates and diffuses. This model is commonly known as the pseudo-two-dimensional (P2D) model, because of the two scales involved (electrode particles and cell scales) or as the Doyle, Fuller \& Newman (DFN) model. In this paper we will refer to it as the DFN model. This model has been since the backbone of lithium ion battery modelling and, more generally, of porous electrode batteries. As most of the modelling efforts on porous electrode batteries have been driven by lithium ion batteries, most of the references and notation in this article refer to that particular application, despite that the results presented are more general.

In the models for porous electrode batteries, the equations derived from the thermodynamics are at the microscale. This means that they need to be solved in a very complex geometry in order to capture the porous electrode, which requires a lot of computational power. For this reason effective models, such as the DFN model, have been popular: they are posed at the macroscale, and therefore are computationally cheaper, but still retain most features of the microstructure and the microscale dynamics. One way to obtain effective macroscale equations from the microscale equations is using the volume averaging method, like in \cite{Fuller1994,Moyles2019asymptotic,Plett2015}. This method provides good results, but it requires to either define the effective model ad hoc \cite{Fuller1994} or to neglect some terms in the averaging \cite{Plett2015}. In both cases, the parameters of the macroscale model need to be defined as effective parameters and, therefore, fitted from data. This means that the macroscale parameters cannot be directly and systematically derived from the microstructure, and has led to the use of empirical correlations which are still a subject of debate in the field, such as the Bruggeman correlation \cite{Tjaden2016}. For more details on volume averaging we refer the reader to \cite{Whitaker1999}.

Another approach to upscale the microscale equations is to use asymptotic homogenisation. The main difference between volume averaging and homogenisation is that the latter allows the effective parameters from the microstructure to be derived in a systematic manner. The key idea of asymptotic homogenisation is to assume that the material is composed of a periodic structure of a size much smaller compared to that of the material. Then, one can exploit the disparity of scales and define space variables at each level that can be treated as independent, so all the differential operators can be split into operators at each scale and a small parameter (ratio of length scales) appears. By performing an asymptotic expansion in this small parameter an equation at the large scale can be determined, which accounts for the effects occurring at the small scale. Homogenisation is a well-known technique: the technical details can be found in the handbooks of the subject \cite{Bensoussan2011,Pavliotis2008} and a detailed comparison between volume averaging and asymptotic homogenisation is provided in \cite{Davit2013}. 

In our case, we have a problem in which there is transport of mass around the particles (i.e. the microstructure), with a chemical reaction at the surface of the particles and diffusion inside them. This type of problem appears in many different applications, apart from porous electrode batteries, such as filtration \cite{Dalwadi2016,Dalwadi2015}, biology \cite{Dalwadi2018}, metallurgical furnaces \cite{Sloman2019} or even coffee roasting \cite{Sachak-Patwa2019}. In order to capture the diffusion in the particles that form the microstructure, it is necessary to use high-contrast homogenisation as the diffusion coeffient in the particles is much smaller than the diffusion coefficient in the medium surrounding them \cite{Ecker2015ii,Ecker2015i}. High-contrast homogenisation is not as developed as standard homogenisation, and it presents numerous challenges, especially on the analysis side \cite{Cherednichenko2016,VanNoorden2011,Zhikov2000}.

Thermal effects can have a  notable impact on the behaviour of batteries, and therefore they are an important aspect of our model. The thermal model is coupled to the electrochemical model in the following way: the currents in the battery generate heat in several ways, while the temperature of the battery affects the parameters of the electrochemical model. The coupled thermal electrochemical model can then be upscaled from the microscale equations from a few different ways, similarly to the electrochemical model. The most common approaches in the literature are posing the model ad hoc \cite{Mei2019} or using volume averaging \cite{Gu2000,Plett2015}, which present the same advantages and disadvantages discussed for the purely electrochemical model. The goal of this paper is to upscale and derive the coupled thermal-electrochemical model systematically, using asymptotic homogenisation, and including some features that extend the models present in the literature. First, we keep the relation between the microstructure and the effective parameters, and also consider a more general microstructure than the single particle at each point considered in the classic DFN model. We also retain all the features in the electrochemistry of the DFN model, in particular diffusion in the particles, which is important in the behaviour of the batteries, especially during the relaxation period after the current is switched off. Finally, we include the double layer capacitance effects in the kinetics between the electrode and the electrolyte, similarly to the model in \cite{Moyles2019asymptotic}.

Homogenisation was first used to derive an effective electrochemical model for porous electrode batteries by Ciucci and Lai in \cite{Ciucci2011}. Assuming that the microstructure is formed by packed spheres they derive the DFN model from the microscale equations. They use high-contrast homogenisation so diffusion in the particles is retained.

A similar approach was taken by Richardson et al. in \cite{Richardson2012}. Here, the authors homogenise the electrochemical microscale equations to obtain effective macroscale equations and use the following assumptions: fast diffusion in the electrode particles, high electronic conductivity in the electrodes and dilute electrolyte. They present a very detailed analysis of the homogenisation and then perform an asymptotic analysis of the effective model to obtain analytical solutions.

In \cite{Arunachalam2017,Arunachalam2015}, Arunachalam et al. perform asymptotic homogenisation of the electrochemical microscale model. The authors consider the diffusivity in the particles to be the same order of magnitude as the diffusivity in the electrolyte. Therefore, the effective model includes lithium diffusion over the whole thickness of the electrode, rather than within the particle. In \cite{Arunachalam2015} the authors also provide an analysis of the region in the parameter space in which the DFN model is valid.

Focusing on another aspect within batteries, Hennessy and Moyles \cite{Hennessy2019} used homogenisation to derive the battery heat equation for a double coated electrode from the cell heat equation. The thermal properties of the cell problem are taken to be scalars and the connection to the microscale bulk properties is not considered.

The outline of this paper is as follows. In Section~\ref{sec:homogenised_model} we provide the homogenised dimensionless model for thermal and electrochemical behaviour of porous electrode batteries. The microscale equations which are the starting point of the derivation of the homogenised model are presented in Section~\ref{sec:microscale} and the details of the homogenisation procedure are given in Section~\ref{sec:homogenisation}. In Section~\ref{sec:comparison_DFN} we compare our homogenised model with the widely used DFN model. Finally, in Section~\ref{sec:discussion} we discuss the results.

\section{Dimensionless homogenised model}\label{sec:homogenised_model}
We use asymptotic homogenisation to derive a thermal-electrochemical model for porous electrode batteries both at cell and battery levels, as this method allows us to derive macroscale equations from the microscale ones in a systematic way. In this section the effective model is presented. This model can be seen as a generalised version of the Doyle, Fuller \& Newman (DFN) model \cite{Fuller1994} and the physical laws underpinning it are discussed at the beginning of Section~\ref{sec:microscale}. Notice that, even though we use lithium-ion battery terminology in the article, the model presented here can be applied to other types of porous electrode batteries.

The variables of the problem are the concentration of lithium in the electrode $c_\mrs$ and ions in the electrolyte $c_\mre$, the electric potentials $\Phi_\mrs$ and $\Phi_\mre$, and the temperature $T$. The subscripts $\mrs$ and $\mre$ distinguish the quantities in the solid electrode and the electrolyte, respectively. For convenience, the currents $\vb*{i}_\mrs$ and $\vb*{i}_\mre$ in the electrode and the electrolyte are defined. These are quantities derived from the concentrations and potentials.

\begin{figure}
	\centering
	\includegraphics[width=\textwidth]{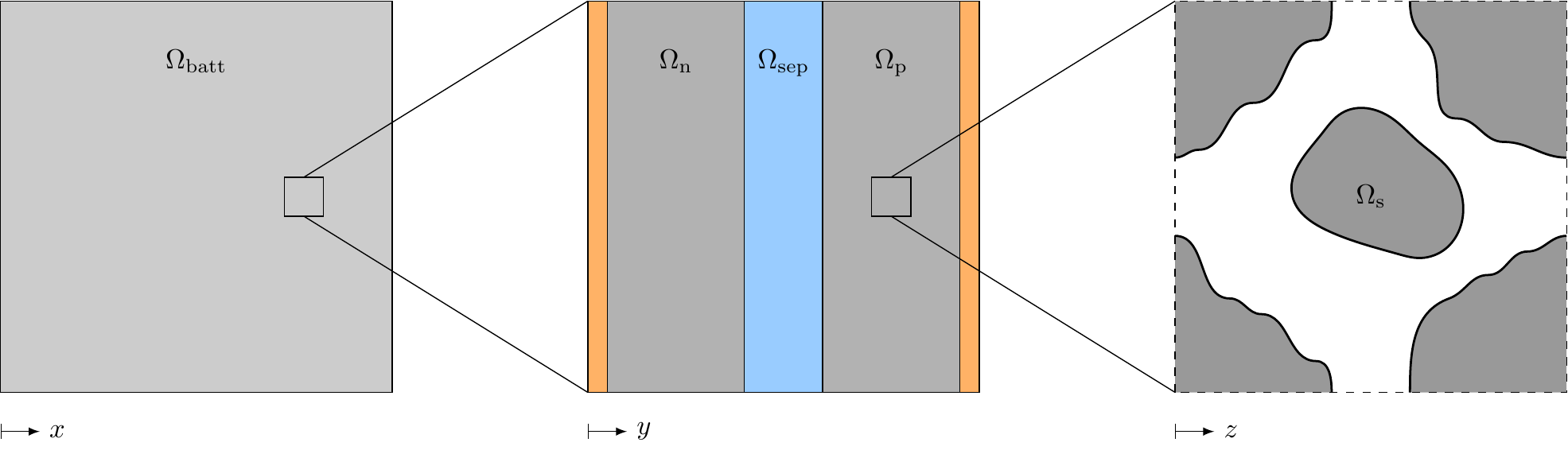}
	\caption{Sketch of the geometry at the three length scales that are involved in the effective model: battery, cell and particles. We label as well the regions at each scale. At the macroscale, where the macroscale heat equation \eqref{eq:heat_eqn_macroscale} is defined, the domain is the battery $\Omega_{\mathrm{batt}}$, which is composed of several cells. At the mesoscale, where the electrochemical model \eqref{eq:cons_charge_meso}-\eqref{eq:electrolyte_eqns} and the cell thermal model \eqref{eq:heat_eqn_mesoscale} are defined, the domains are the negative electrode $\Omega_\mathrm{n}$, the separator $\Omega_{\mathrm{sep}}$, and the positive electrode $\Omega_\mathrm{p}$ (apart from the collectors which have not been labelled as we do not consider them in the problem). The cell is defined as the union of the three parts so $\Omega_{\mathrm{cell}} = \Omega_\mathrm{n} \cup \Omega_{\mathrm{sep}} \cup \Omega_\mathrm{p}$. At the microscale, where the diffusion in the particles \eqref{eq:cons_mass_s_homog} is defined, the domain is the representative microstructure of the porous structure $\Omega_\mathrm{s}$.}
	\label{fig:multiscale}
\end{figure}

As shown in Figure \ref{fig:multiscale}, the effective model accounts for phenomena occurring at three different scales. At the microscale, denoted by $z$, there are the electrode particles $\Omega_\mrs$. The mesoscale, denoted by $y$, represents a cell, which is composed of the negative electrode $\Omega_\mathrm{n}$, the separator $\Omega_\mathrm{sep}$, and the positive electrode $\Omega_\mathrm{p}$. The boundary of the domain needs to be separated into the part in contact with the current collectors (denoted by $\partial \Omega_\mathrm{n}^\mathrm{collector}$ and $\partial \Omega_\mathrm{p}^\mathrm{collector}$) and the rest of the boundary, as different conditions apply to each one. At each point of the cell there is the solid electrode and the electrolyte and thus different variables are used to measure them. The cell domain, defined as the union of the three parts, is $\Omega_\mathrm{cell}$. Finally, there is the macroscale, denoted by $x$, which is the battery, composed of several cells and represented by $\Omega_\mathrm{batt}$. In the differential operator $\nabla$ subscripts $x$, $y$, and $z$ are used to denote at which scale the operator is applied. Similarly, we use these subscripts in the volume and area integration variables to denote the scale.

Then, the homogenised dimensionless problem is the following. For the concentration of lithium in the electrodes the problem is defined at the microscale domain. It reads
\begin{subequations}\label{eq:cons_mass_s_homog}
	\begin{align}
	\dt c_\mathrm{s} &= \nabla_z \cdot \left(D_\mathrm{s} \nabla_z c_\mathrm{s} \right), & \text{ in } \Omega_\mathrm{s},\\
	-D_\mathrm{s} \nabla_z c_\mathrm{s} \cdot \vb*{n}_\mathrm{s} &= G  \left(g + C \dt\left(\Phi_\mathrm{s} - \Phi_\mathrm{e} \right)  \right), & \text{ at } \partial\Omega_\mathrm{s}^{\mathrm{in}},
	\end{align}
	where $D_\mrs$ is the diffusion coefficient of lithium in the electrode (which may depend on the concentration itself, space variables and temperature), $G$ is the ratio between applied and exchange current, and $C$ is the double layer capacitance. The exchange current $g$ and the overpotential $\eta$ are defined as
	\begin{align}
	g &= c_\mathrm{s}^{\beta} \left(1 - \frac{c_\mathrm{s}}{c_\mathrm{s}^{\max}} \right)^{1-\beta} c_{\mathrm{e}}^{1-\beta} \left[\exp\left( (1-\beta) \lambda \frac{\eta}{1 + \gamma T} \right) - \exp\left( - \beta \lambda \frac{\eta}{1 + \gamma T} \right) \right],\label{eq:def_g}\\ 
	\eta &= \Phi_\mathrm{s}- \Phi_\mathrm{e} - U_{\mathrm{ocp}}\left(c_\mathrm{s}\right),\label{eq:def_eta}
	\end{align}
	 where $\beta$ is the charge-transfer coefficient of the interacalation reaction, $c_\mrs^{\max}$ is the maximum concentration of the electrode, $\lambda$ is the ratio between the typical and thermal potentials, $\gamma$ is the ratio between temperature variation and reference temperature, and $U_{\mathrm{ocp}}$ is the open circuit potential as a function of the electrode lithium concentration at the interface. All these parameters are dimensionless quantities, and more details about the non-dimensionalisation are provided in Appendix~\ref{sec:nondim}. In addition, $c_s$ must be periodic in $z \in \mathbb{R}^3$. Even though here we consider the electrode particles to be isotropic, the same model would apply to anisotropic materials just taking the diffusion coefficients $D_\mrs$ to be tensors.
\end{subequations}

At the mesoscale, the conservation of charge equations are posed in each electrode separately
\begin{subequations}\label{eq:cons_charge_meso}
	\begin{align}
	\nabla_y \cdot \vb*{i}_\mathrm{s} &= -J, & \text{ in } \Omega_\mathrm{p},\\
	\vb*{i}_\mathrm{s} \cdot \vb*{n}_\mathrm{p} &= - i_\mathrm{app}(y,t), & \text{ at } \partial\Omega_\mathrm{p}^{\mathrm{collector}},\label{eq:cons_charge_meso_b}\\
	\vb*{i}_\mathrm{s} \cdot \vb*{n}_\mathrm{p} &= 0, & \text{ at } \partial \Omega_\mathrm{p} \setminus \partial \Omega_\mathrm{p}^{\mathrm{collector}},
	\end{align}
	with the current $\vb*{i}_\mrs$ and the exchange current per unit of volume $J$ defined as
	\begin{align}
	\vb*{i}_\mathrm{s} &= -\mathcal{S} \nabla_y \Phi_\mathrm{s},\label{eq:def_is}\\
	J &= \frac{1}{|\Omega|} \int_{\partial \Omega_\mathrm{s}^{\mathrm{in}}} G \left(g + C \dt \left(\Phi_\mathrm{s} - \Phi_\mathrm{e} \right) \right) \dd A_z,\label{eq:def_J}
	\end{align}
\end{subequations}
In the previous equations, $i_\mathrm{app}$ is the current density applied to the battery (defined to be positive during discharge), and $\mathcal{S}$ is the mesoscale electronic conductivity tensor in the electrode, which may vary in space and depend on temperature. The domain $\partial \Omega_{\mrs}^\mathrm{in}$ is the electrode-electrolyte contact surface as shown in Figure \ref{fig:microscale_domain}. Because in this model we do not consider the current collectors, we assume that the applied current density $i_\mathrm{app}$ is known. However, in most of the cell configurations, in order to determine $i_\mathrm{app}$ an additional problem at the macroscale for the specific geometry of the cell. This has been addressed in the literature mostly from a numerical point of view \cite{Gerver2011,Guo2013,Kosch2018}. The incorporation of the current collectors into this homogenisation analysis is an area of future work.

Notice that two instances of problems \eqref{eq:cons_mass_s_homog} and \eqref{eq:cons_charge_meso} are needed to describe a cell, one for each electrode. The problem in \eqref{eq:cons_charge_meso} is for the positive electrode, while the problem in the negative electrode is the same but with the boundary condition \eqref{eq:cons_charge_meso_b} having the opposite sign. In each electrode the parameters can take different values, but they are constant in each electrode. Also, the microstructure, defined by $\Omega_s$, is different in each electrode and the concentration fields $c_\mathrm{s}$ may have significantly different behaviours in each electrode.

The electrolyte problem can be posed across the electrodes and the separator, and it is given by
\begin{subequations}\label{eq:electrolyte_eqns}
	\begin{align}
	\varphi_\mathrm{e} \dt c_\mathrm{e} - \nabla_y \cdot \left(\mathcal{D}_\mathrm{L} \nabla_y c_\mathrm{e} + \lambda \mathcal{M}_\mathrm{L} c_\mathrm{e} \nabla_y \Phi_\mathrm{e} \right) &= J, & \text{ in } \Omega_\mathrm{cell},\label{l_conc}\\
	\nabla_y \cdot \vb*{i}_\mre &= J, & \text{ in } \Omega_\mathrm{cell},\label{c_cons}\\
	- \left(\mathcal{D}_\mathrm{L} \nabla_y c_\mathrm{e} + \lambda \mathcal{M}_\mathrm{L} c_\mathrm{e} \nabla_y \Phi_\mathrm{e} \right) \cdot \vb*{n}_\mathrm{cell} &= 0, & \text{ at } \partial \Omega_\mathrm{cell},\\
	\vb*{i}_\mre \cdot \vb*{n}_\mathrm{cell} &= 0, & \text{ at } \partial \Omega_\mathrm{cell},
	\end{align}
	with the current in the electrolyte $\vb*{i}_\mre$ defined as
	\begin{align}
	\vb*{i}_\mre &= - \left((\mathcal{D}_\mathrm{L} - \mathcal{D}_\mathrm{A}) \nabla_y c_\mathrm{e} + \lambda (\mathcal{M}_\mathrm{L} + \mathcal{M}_\mathrm{A}) c_\mathrm{e} \nabla_y \Phi_\mathrm{e} \right).\label{eq:def_ie}
	\end{align}
\end{subequations}
The parameters $\mathcal{D}_\mathrm{L}$ and $\mathcal{D}_\mathrm{A}$ are the mesoscale diffusivity tensors for lithium ions and anions in the electrolyte respectively (that is, accounting for the microstructure), $\mathcal{M}_\mathrm{L}$ and $\mathcal{M}_\mathrm{A}$ are the mesoscale ion mobilities (which are tensors too), and $\varphi_\mre$ is the volume fraction occupied by the electrolyte. As we show during the homogenisation process, $\mathcal{D}_\mathrm{L}$ and $\mathcal{D}_\mathrm{A}$ must be multiples of each other as any anisotropy arising in the electrolyte can be caused only by the geometry, which is the same for both types of ions, and the same applies to the pair $\mathcal{M}_\mathrm{L}$ and $\mathcal{M}_\mathrm{A}$. Both diffusivities and mobilities may depend on the ion concentration in the electrolyte and temperature. All these parameters, even if they are constant in each part (electrodes and separator) like the electrolyte volume fraction, have different values in each part. In particular, $J$ is equal to zero in the separator as there is no reaction taking place. Notice as well that the electrolyte equations only consider the concentration of lithium ions because electroneutrality has been assumed in the electrolyte, and therefore the concentration of cations and anions must be the identical.

Because this is a homogenised model, we assume that $D_\mrs$, $\mathcal{D}_\mathrm{L}$ and $\mathcal{D}_\mathrm{A}$ are all of the same order of magnitude. Even though diffusion in the electrode is much slower than in the electrolyte, this effect has already been captured by the fact that diffusion in the electrode occurs at a much smaller length scale. In particular, $D_\mrs$ is the same as in the microscale model in Section~\ref{sec:microscale}, while $\mathcal{D}_\mathrm{L} = D_\mathrm{L} \mathcal{B}$ and $\mathcal{D}_\mathrm{A} = D_\mathrm{A} \mathcal{B}$, where $D_\mathrm{L}$ and $D_\mathrm{A}$ are the microscale diffusivities and $\mathcal{B}$ is the tensor accounting for the geometry of the porous material as defined in Section~\ref{sec:properties_tensors}.

The mesoscale heat equation is given by
\begin{subequations}\label{eq:heat_eqn_mesoscale}
	\begin{align}
	\theta \dt T &= \nabla_y \cdot \left(\mathcal{K} \nabla_y T \right) + Q,
	\end{align}
	where $\theta$ is the volumetric heat capacity and $\mathcal{K}$ is the thermal conductivity tensor. The heat source term $Q$ accounts for four different heat generation mechanisms: Joule heating in the electrode, Joule heating in the electrolyte, irreversible reaction heating and reversible reaction heating. These terms can be written as
	\begin{align}
	Q_\mathrm{s} &= - \lambda \vb*{i}_\mathrm{s} \cdot \left(\mathcal{Q}_\mathrm{s} \nabla_y \Phi_\mathrm{s} \right),\label{eq:def_Qs}\\
	Q_\mathrm{e} &= - \lambda \vb*{i}_\mathrm{e} \cdot \left(\mathcal{Q}_\mathrm{e} \nabla_y \Phi_\mathrm{e} \right),\label{eq:def_Qe}\\
	Q_{\mathrm{irr}} &= \frac{1}{|\Omega |} \int_{\partial \Omega_\mathrm{s}^{\mathrm{in}}} \lambda G g \eta \dd A_z,\\
	Q_{\mathrm{rev}} &= \frac{1}{|\Omega |} \int_{\partial \Omega_\mathrm{s}^{\mathrm{in}}} \lambda G g \Pi \dd A_z,
	\end{align}
	where $\mathcal{Q}_\mathrm{s}$ and $\mathcal{Q}_\mathrm{e}$ are tensors that account for the microstructure effects in the heat generation, and $\Pi$ is the Peltier term. The Peltier term is defined as
	\begin{equation}
	    \Pi = T \pdv{U_\mathrm{ocp}}{T},
	\end{equation}
	so it is temperature dependent. However, given that in many practical applications the $U_\mathrm{ocp}$ is provided from experimental data (and so is its derivative with respect to temperature), we treat the Peltier term as a parameter function of the model.
	
	Then, the heat generation term is defined as
	\begin{equation}\label{eq:def_Q}
	Q = \begin{cases}
	Q_\mathrm{s} + Q_\mathrm{e} + Q_{\mathrm{irr}} + Q_{\mathrm{rev}}, & \text{ in } \Omega_\mathrm{p} \text{ and } \Omega_\mathrm{n}, \\
	Q_\mathrm{e}, & \text{ in } \Omega_\mathrm{sep}.
	\end{cases}
	\end{equation}
\end{subequations}
The boundary conditions are not specified at this point as they depend on the problem of interest. To study a single cell, heat exchange conditions at the boundary could be used. If, instead, this equation is to be homogenised to obtain the battery heat equation, then periodic boundary conditions are required.

In a similar way, the heat equation at the battery level can be defined as
\begin{subequations}\label{eq:heat_eqn_macroscale}
	\begin{align}
	\theta_{\mathrm{batt}} \dt T &= \nabla_x \cdot \left(\mathcal{K}_{\mathrm{batt}} \nabla_x T \right) + Q_{\mathrm{batt}}, & \text{ in } \Omega_\mathrm{batt},
	\end{align}
	with suitable boundary conditions, where $\theta_{\mathrm{batt}}$ is the average volumetric heat capacity of a cell, and $\mathcal{K}_{\mathrm{batt}}$ is the thermal conductivity tensor of the battery. The heat source term $Q_{\mathrm{batt}}$ accounts for the heat generation in each cell and it is defined as
	\begin{equation}\label{eq:def_Q_batt}
	\begin{aligned}
	Q_{\mathrm{batt}} &= \frac{1}{|\Omega_{\mathrm{cell}}|} \left(\int_{\Omega_p} \left(Q_{\mathrm{s}} + Q_{\mathrm{irr}} + Q_{\mathrm{rev}} \right) \dd V_y \right. \\
	& \quad \left. + \int_{\Omega_n} \left( Q_{\mathrm{s}} + Q_{\mathrm{irr}} + Q_{\mathrm{rev}} \right) \dd V_y + \int_{\Omega_{\mathrm{cell}}} Q_e \dd V_y \right)\\
	&= \frac{1}{|\Omega_{\mathrm{cell}}|} \int_{\Omega_{\mathrm{cell}}} Q \dd V_y.
	\end{aligned}
	\end{equation}
\end{subequations}
An extra term could be added to $Q_\mathrm{batt}$ to account for the heat generated in the current collectors, but for simplicity of the homogenisation problem we have not considered it.

All the tensors that appear in the homogenised model can be calculated from material properties and microstructure, as detailed in Section~\ref{sec:properties_tensors}.

\section{Dimensionless microscale model}\label{sec:microscale}
We now consider the dimensionless microscale model, from which we derive the effective model. The electrochemical model accounts for conservation of mass and charge in both the electrodes and the electrolyte, with Butler-Volmer kinetics for the intercalation reaction. The transport of lithium in the electrodes is by diffusion only, while in the electrolyte there is diffusion and migration due to the electric field. We use a dilute electrolyte model, and thus use Nernst-Planck equations, but analogous results can be found for concentrated electrolytes (see \cite{Arunachalam2015,Ciucci2011}). We assume as well that charge transport in the electrodes follows Ohm's law. The thermal model imposes conservation of energy in both electrode and electrolyte accounting only by diffusion effects, with a heat source term at the interface between them due to the chemical reaction. For a complete discussion of the microscale equations we refer the reader to the books \cite{Newman2004,Plett2015}.

The variables for the microscale problem are still the concentration of lithium in the electrodes $c_\mrs$ and lithium ions in the electrolyte $c_\mre$, the potentials $\Phi_\mrs$ and $\Phi_\mre$, and the temperature $T$. To be rigorous, these variables are not the same as those defined in Section~\ref{sec:homogenised_model}, which are the leading order term in the asymptotic expansions of the microscale variables. However, to keep the notation simple, we do not use any symbol to distinguish them because for the rest of the paper we will refer to the microscale variables and their asymptotic expansions. For fluxes, on the other hand, we need to distinguish the homogenised ones from the bulk ones observed at the microscale, therefore we use tilde for the fluxes at the microscale (i.e. $\tilde{\vb*{N}}_\mathrm{s}$, $\tilde{\vb*{i}}_\mathrm{s}$ and $\tilde{\vb*{K}}$).

\begin{figure}
	\centering
	\includegraphics[scale = 0.75]{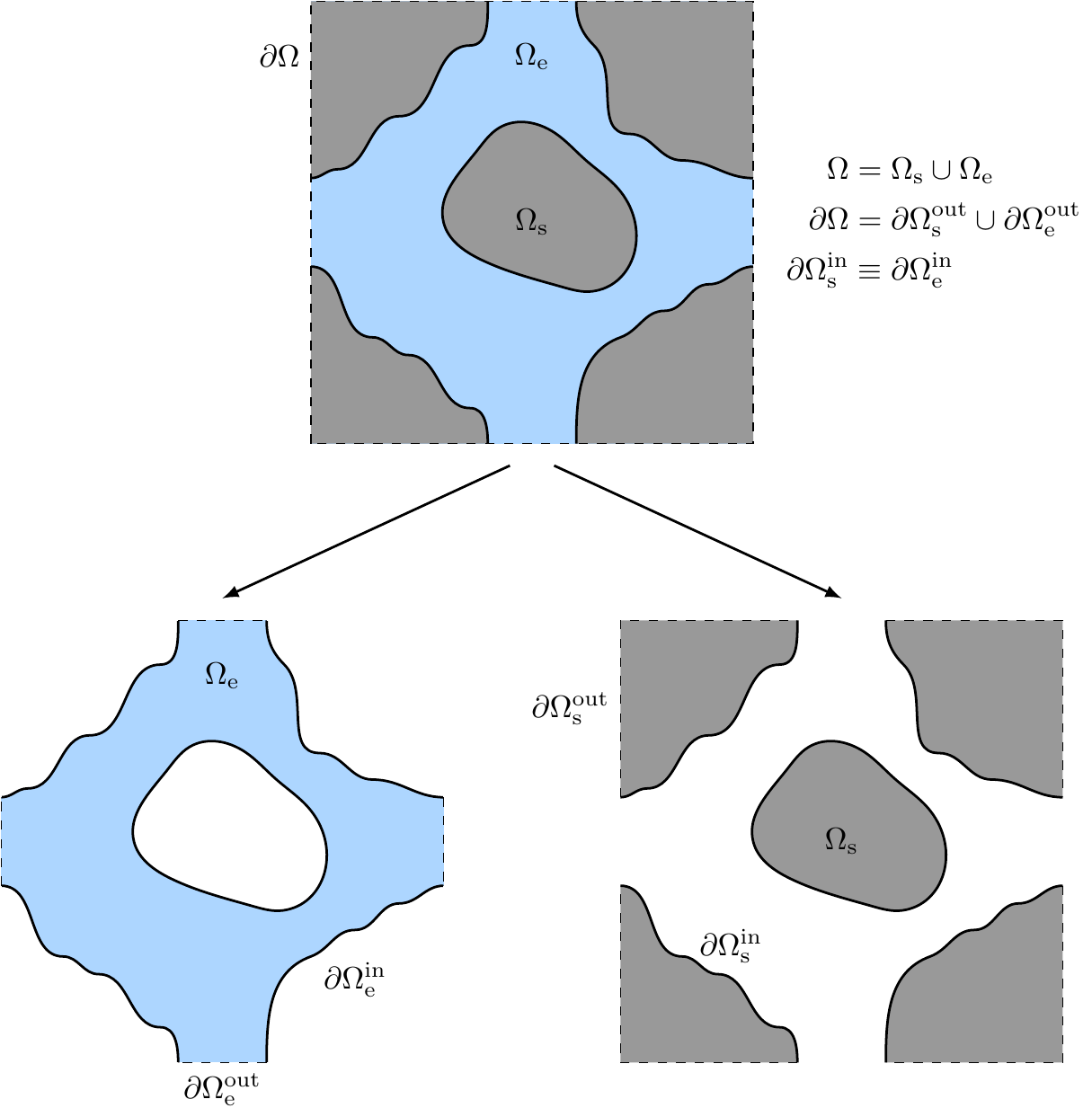}
	\caption{Schematic of the domain definition at the microscale. The microscale periodic cell $\Omega$ is composed of the electrolyte domain $\Omega_\mathrm{e}$ and the solid electrode domain $\Omega_\mathrm{s}$. For the homogenisation problem we define $\partial\Omega$ as the boundary of $\Omega$, which is composed by $\partial\Omega_\mathrm{s}^{\mathrm{out}}$ and $\partial\Omega_\mathrm{e}^{\mathrm{out}}$ depending on whether the boundary is at the solid or the electrolyte domain. The interface between electrolyte and solid is defined by both $\partial\Omega_\mathrm{e}^{\mathrm{in}}$ and $\partial\Omega_\mathrm{s}^{\mathrm{in}}$.}
	\label{fig:microscale_domain}
\end{figure}

The microscale equations are defined in a domain $\Omega$, which is divided into the electrode particles $\Omega_\mrs$ and the electrolyte $\Omega_\mre$. In the homogenisation process we consider the electrode and electrolyte separately (except for the thermal model), therefore we need to carefully define the notation in each domain. The thermal problem is defined as a single equation in the joint domain $\Omega$ because the thermal problem is the same in both electrode and electrolyte, just with different parameters. A detailed sketch of each of the subdomains and their boundaries is shown in Figure \ref{fig:microscale_domain}. A key assumption for the homogenisation is that the microstructure repeats periodically, therefore a finite domain that is representative of the geometry can be considered. The length scale of $\Omega$ is much smaller than the length scale of the porous electrode. Therefore, the ratio between length scales $\delta$, which is a small parameter, arises in the equations and can be used in the homogenisation from microscale to mesoscale.

The details on the non-dimensionalisation of the microscale model are provided in Appendix~\ref{sec:nondim}. The scalings are chosen so all the parameters in the dimensionless model are of $\order{1}$ except for $\delta$ which is small. The dimensionless equations for the solid phase are
\begin{subequations}\label{eq:electrode_micro}
	\begin{align}
	\dt c_\mrs + \nabla \cdot \tilde{\vb*{N}}_\mrs &= 0, & \text{ in } \Omega_\mrs,\\
	\nabla \cdot \tilde{\vb*{i}}_\mrs &= 0, & \text{ in } \Omega_\mrs,\\
	\tilde{\vb*{N}}_\mrs \cdot \vb*{n}_\mrs &= \delta G  \left(g + C \dt\left(\Phi_\mrs - \Phi_\mre \right)  \right), & \text{ at } \partial\Omega_\mrs^{\mathrm{in}},\label{eq:electrode_micro_c}\\
	\tilde{\vb*{i}}_\mrs \cdot \vb*{n}_\mrs &= \delta G  \left(g + C \dt\left(\Phi_\mrs - \Phi_\mre \right) \right), & \text{ at } \partial\Omega_\mrs^{\mathrm{in}},\label{eq:electrode_micro_d}\\
	& \text{periodic}, & \text{ at } \partial\Omega_\mrs^{\mathrm{out}},
	\end{align}
	with
	\begin{align}
	\tilde{\vb*{N}}_\mrs &= -\delta^2 D_\mrs \nabla c_\mrs,\\
	\tilde{\vb*{i}}_\mrs &= - \sigma_\mrs \nabla \Phi_\mrs,
	\end{align}
	and
	\begin{equation}
	g = c_\mrs^\beta \left(1 - \frac{c_\mrs}{c_\mrs^{\max}} \right)^{1-\beta} c_\mre^{1-\beta} \left[\exp\left( (1-\beta) \lambda \frac{\eta}{1 + \gamma T} \right) - \exp\left( - \beta \lambda \frac{\eta}{1 + \gamma T} \right) \right],
	\end{equation}
\end{subequations}
where the only new parameters are the ratio between length scales at micro and mesoscale $\delta$, and the microscale electronic conductivity $\sigma_\mrs$. The latter may depend on temperature and the spacial variables in order to account for inhomogeneities in the electrode material. Diffusion of lithium in the solid is much slower than diffusion of ions in the electrolyte, therefore we consider the limit in which the diffusivity in the solid is of $\order{\delta^2}$. As pointed out in \cite{Sloman2019}, this particular limit allows diffusion to happen only at the microscale, as we expect in a porous electrode battery.

The equations for the electrolyte are
\begin{subequations}\label{eq:electrolyte_micro}
	\begin{align}
	\dt c_\mre + \nabla \cdot \tilde{\vb*{N}}_\mre &= 0, & \text{ in } \Omega_\mre,\\
	\nabla \cdot \tilde{\vb*{i}}_\mre &= 0, & \text{ in } \Omega_\mre,\\
	\tilde{\vb*{N}}_\mre \cdot \vb*{n}_\mre &= -\delta G  \left(g + C \dt\left(\Phi_\mrs - \Phi_\mre \right) \right), &  \text{ at } \partial\Omega_\mre^{\mathrm{in}},\label{eq:electrolyte_micro_c}\\
	\tilde{\vb*{i}}_\mre \cdot \vb*{n}_\mre &= -\delta G  \left(g + C \dt\left(\Phi_\mrs - \Phi_\mre \right) \right), &  \text{ at } \partial\Omega_\mre^{\mathrm{in}},\label{eq:electrolyte_micro_d}\\
	& \text{periodic}, & \text{ at } \partial\Omega_\mre^{\mathrm{out}},
	\end{align}
	with
	\begin{align}
	\tilde{\vb*{N}}_\mre &= -\left(D_\mathrm{L} \nabla c_\mre + \lambda \mu_\mathrm{L} c_\mre \nabla \Phi_\mre \right),\\
	\tilde{\vb*{i}}_\mre &= -\left((D_\mathrm{L} - D_\mathrm{A}) \nabla c_\mre + \lambda (\mu_\mathrm{L} + \mu_\mathrm{A}) c_\mre \nabla \Phi_\mre \right).
	\end{align}
\end{subequations}
The new parameters here are $D_\mathrm{L}$ and $D_\mathrm{A}$ which are the diffusion coefficients of lithium ions and anions in the electrolyte, and $\mu_\mathrm{L}$ and $\mu_\mathrm{A}$ which are the ion mobilities. These are scalars as they are bulk parameters and do not account for the porous structure of the medium. We assume that diffusivities and mobilities of ions are homogeneous in space and do not depend on the concentration of ions or the voltage. However, they may depend on temperature. Notice the $\delta$ factor in the fluxes \eqref{eq:electrode_micro_c}-\eqref{eq:electrode_micro_d} and \eqref{eq:electrolyte_micro_c}-\eqref{eq:electrolyte_micro_d}. This scaling arises from the fact that the surface area of the particles is of $\order{\delta^{-1}}$, and given that the total exchange current over each electrode is of $\order{1}$ this implies that the exchange currents (and ion fluxes) must be of $\order{\delta}$. To pose the electrolyte microscale equations we have assumed electroneutrality, which is true except in a thin layer at the electrode-electrolyte interface. The size of this layer is defined by the Debye length, which is of the order of nanometres. Therefore, even at the microscale (which is of the order of micrometers) the electroneutrality assumption is reasonable.

The dimensionless thermal model is
\begin{subequations}\label{eq:temp_micro}
	\begin{align}
	\rho c_p \dt T + \nabla \cdot \tilde{\vb*{K}} &= - \lambda \tilde{\vb*{i}} \cdot \nabla \Phi, & \quad \text{ in } \Omega,\\
	\left(\tilde{\vb*{K}}_\mrs - \tilde{\vb*{K}}_\mre \right) \cdot \vb*{n}_\mre &= \delta \lambda G (g \eta + g \Pi), & \quad \text{ in } \partial \Omega^{\mathrm{in}}_\mre,\\
	& \text{periodic}, & \quad \text{ at } \partial \Omega,
	\end{align}
	where
	\begin{equation}
	\tilde{\vb*{K}} = -k \nabla T,
	\end{equation}
	and $\rho$ is the density, $c_p$ the specific heat capacity, and $k$ the thermal conductivity. All these parameters depend on $z$ as they are different in each material, and they can vary within each material too. We also have that $\tilde{\vb*{i}}$ and $\Phi$ correspond to $\tilde{\vb*{i}}_\mre$ and $\Phi_\mre$, or $\tilde{\vb*{i}}_\mrs$ and $\Phi_\mrs$ depending on whether the point in the domain is in $\Omega_\mre$ or $\Omega_\mrs$. We use $\tilde{\vb*{K}}_\mrs$ to denote the flux on the electrode side of the interface and $\tilde{\vb*{K}}_\mre$ to denote the flux on the electrolyte side.
\end{subequations}

These microscale equations account for the classic conservation of mass, charge and energy laws, so they are physically consistent (see \cite{Newman2004,Plett2015} for details). However, given the complexity of the porous structure and the multiple scales involved in the problem, solving this model numerically is very challenging. The motivation to derive the homogenised mesoscale equations is to obtain a model of a lower complexity level that still captures most of the microstructure effects.

\section{Derivation of the effective equations}\label{sec:homogenisation}
We now proceed to homogenise the microscale equations defined in Section~\ref{sec:microscale} in order to derive the effective equations presented in Section~\ref{sec:homogenised_model}. To do so, we need to homogenise the equations at the microscale (i.e. porous material structure) to obtain the mesoscale equations (i.e. cell level). Then, we can assemble the mesoscale equations for a cell and homogenise them to obtain the macroscale equations (i.e. battery level). 

In our problem, we have a particle microstructure of length scale $\ell$ composing a porous material of length scale $L$ which makes a cell. A large number of these cells are put together to make a battery of length scale $N L$, where $N$ is number of cells in the battery. Then, we define the following dimensionless numbers
\begin{equation}
\delta = \frac{\ell}{L} \quad \text{ and } \quad \epsilon = \frac{1}{N},
\end{equation}
and we have that both $\delta,\epsilon\ll 1$. These are the small numbers that we exploit for the homogenisation. In particular, for lithium ion batteries, using as typical values those in \cite{Ecker2015ii,Ecker2015i} and that the radius of a typical cylindrical cells is around 1 cm \cite{Quinn2018}, we estimate values of $\delta$ and $\epsilon$ of the order of $10^{-2}$.

Taking the limit of small $\delta$ and $\epsilon$, we can apply the chain rule to split the differential operators into the multiple scales. Recall, that the variable at the macroscale is $x$, at the mesoscale is $y$, and at the microscale is $z$. Then, when deriving the mesoscale equations, the operator acts on both the meso and microscales, so for a general function $f$ we have
\begin{subequations}\label{eq:splitting_nabla}
	\begin{align}\label{eq:splitting_nabla_delta}
	\nabla f\left(y,z\right) = \nabla f\left(y,\frac{y}{\delta}\right)&= \nabla_y f(y,z) + \frac{1}{\delta} \nabla_z f(y,z).
	\end{align}
	When we assemble the cell model and upscale it to the battery level, the macro and mesoscales are involved, so we have
	\begin{align}\label{eq:splitting_nabla_eps}
	\nabla f\left(x,y\right) = \nabla f\left(x,\frac{x}{\epsilon}\right) &= \nabla_x f(x,y) + \frac{1}{\epsilon} \nabla_y f(x,y),
	\end{align}
\end{subequations}
where the subscripts in $\nabla$ denote with respect to which variable the operator is applied. 

The homogenisation analysis presented here assumes that all the parameters except for $\delta$ or $\epsilon$ are of $\order{1}$. In practise, for the typical lithium-ion chemistries, some of these parameter are big or small but this does not affect the homogenisation analysis. Having defined the different scales and the operators, we can now proceed to homogenise the equations.

\subsection{Conservation of mass in the electrode}
We start by homogenising the conservation of mass in the electrode, which written accounting for the multiple scales, is given by
\begin{subequations}\label{eq:cons_mass_s_multiscale}
	\begin{align}
	\dt c_\mrs + \nabla_y \cdot \tilde{\vb*{N}}_\mrs + \frac{1}{\delta} \nabla_z \cdot \tilde{\vb*{N}}_\mrs &= 0, & \text{ in } \Omega_\mrs,\\
	\tilde{\vb*{N}}_\mrs \cdot \vb*{n}_\mrs &= \delta G  \left(g + C \dt\left(\Phi_\mrs - \Phi_\mre \right)  \right), & \text{ at } \partial\Omega_\mrs^{\mathrm{in}},\\
	& \text{periodic}, & \text{ at } \partial\Omega_\mrs^{\mathrm{out}},
	\end{align}
	with
	\begin{equation}
	\tilde{\vb*{N}}_\mrs = -\delta^2 D_\mrs \left(\nabla_y c_\mrs + \frac{1}{\delta} \nabla_z c_\mrs \right).
	\end{equation}
\end{subequations}
We expand the concentration and the flux as
\begin{subequations}
	\begin{align}
	c_\mrs &= c_{\mrs,0} + \delta c_{\mrs,1} + \delta^2 c_{\mrs,2} + \order{\delta^3},\\
	\tilde{\vb*{N}}_\mrs &= \delta \tilde{\vb*{N}}_{\mrs,1} + \delta^2 \tilde{\vb*{N}}_{\mrs,2} + \order{\delta^3},
	\end{align}    
\end{subequations}
so now we can substitute these expansions into \eqref{eq:cons_mass_s_multiscale} and linearise the problem.

At leading order we find
\begin{subequations}
	\begin{align}
	\dt c_{\mrs,0} + \nabla_{z} \cdot\tilde{\vb*{N}}_{\mrs,1} &= 0, & \text{ in } \Omega_\mrs,\\
	\tilde{\vb*{N}}_{\mrs,1} \cdot \vb*{n}_\mrs &= G  \left(g_0 + C \dt\left(\Phi_{\mrs,0} - \Phi_{\mre,0} \right)  \right), & \text{ at } \partial\Omega_\mrs^{\mathrm{in}},\\
	& \text{periodic}, & \text{ at } \partial\Omega_\mrs^{\mathrm{out}},
	\end{align}
	with
	\begin{equation}
	\tilde{\vb*{N}}_{\mrs,1} = -D_\mrs \nabla_z c_{\mrs,0},
	\end{equation}
\end{subequations}
where $g_0$, $\Phi_{\mrs,0}$ and $\Phi_{\mre,0}$ are the leading order terms in the expansion for small $\delta$ of $g$, $\Phi_\mrs$ and $\Phi_\mre$ respectively. Then, we conclude that the governing equation for the conservation of mass in the electrodes is
\begin{subequations}
	\begin{align}
	\dt c_{\mrs,0} &= \nabla_{z} \cdot \left(D_\mrs \nabla_z c_{\mrs,0}\right), & \text{ in } \Omega_\mrs,\\
	-D_\mrs \nabla_z c_{\mrs,0} \cdot \vb*{n}_\mrs &= G  \left(g_0 + C \dt\left(\Phi_{\mrs,0} - \Phi_{\mre,0} \right)  \right), & \text{ at } \partial\Omega_\mrs^{\mathrm{in}},\\
	& \text{periodic}, & \text{ at } \partial\Omega_\mrs^{\mathrm{out}},
	\end{align}
\end{subequations}
which is the problem stated in \eqref{eq:cons_mass_s_homog}. Notice that this problem is still at the microscale because the diffusion coefficient is of $\order{\delta^2}$ and, therefore, diffusion is so slow that can only be observed at the microscale. Because the homogenisation process does not change the diffusion coefficients for lithium in the electrode particles, the same result holds for anisotropic materials, just replacing the diffusion coefficient by a tensor.

\subsection{Conservation of charge in the electrode}\label{sec:cons_charge_solid}
Next we consider the conservation of charge in the electrode, which is given by
\begin{subequations}\label{eq:cons_charge_s_multiscale}
	\begin{align}
	\nabla_y \cdot \tilde{\vb*{i}}_\mrs + \frac{1}{\delta} \nabla_z \cdot \tilde{\vb*{i}}_\mrs &= 0, & \text{ in } \Omega_\mrs,\\
	\tilde{\vb*{i}}_\mrs \cdot \vb*{n}_\mrs &= \delta G  \left(g + C \dt\left(\Phi_\mrs - \Phi_\mre \right) \right), & \text{ at } \partial\Omega_\mrs^{\mathrm{in}},\\
	& \text{periodic}, & \text{ at } \partial\Omega_\mrs^{\mathrm{out}},
	\end{align}
	with
	\begin{align}\label{eq:def_is_multiscale}
	\tilde{\vb*{i}}_\mrs &= -\sigma_\mrs \left(\nabla_y \Phi_\mrs + \frac{1}{\delta} \nabla_z \Phi_\mrs \right).
	\end{align}
\end{subequations}
We expand the potential and the current as
\begin{subequations}
	\begin{align}
	\Phi_\mrs &= \Phi_{\mrs,0} + \delta \Phi_{\mrs,1} + \delta^2 \Phi_{\mrs,2} + \order{\delta^3},\\
	\tilde{\vb*{i}}_\mrs &= \delta^{-1} \tilde{\vb*{i}}_{\mrs,-1} + \tilde{\vb*{i}}_{\mrs,0} + \delta \tilde{\vb*{i}}_{\mrs,1} + \order{\delta^2}.
	\end{align}    
\end{subequations}
Notice that the expansion for current starts at $\delta^{-1}$ because of the definition \eqref{eq:def_is_multiscale}. Substituting these expansions into \eqref{eq:cons_charge_s_multiscale}, multiplying by a power of $\delta$ so the leading order term is $\order{1}$, and expanding for small $\delta$ we find the following problems. 

\subsubsection{$O(1)$ problem}\label{sec:cons_mass_s_O1}
At leading order we have the problem
\begin{subequations}\label{eq:cons_mass_s_O1}
	\begin{align}
	\nabla_z \cdot \tilde{\vb*{i}}_{\mrs,-1} &= 0, & \text{ in } \Omega_\mrs,\label{eq:cons_mass_s_O1_a}\\
	\tilde{\vb*{i}}_{\mrs,-1} \cdot \vb*{n}_\mrs &= 0, & \text{ at } \partial\Omega_\mrs^{\mathrm{in}},\label{eq:cons_mass_s_O1_b}\\
	& \text{periodic}, & \text{ at } \partial\Omega_\mrs^{\mathrm{out}},\label{eq:cons_mass_s_O1_c}
	\end{align}
	with
	\begin{align}
	\tilde{\vb*{i}}_{\mrs,-1} &= - \sigma_\mrs \nabla_z \Phi_{\mrs,0}.
	\end{align}    
\end{subequations}
The first step in the homogenisation process is to show that $\Phi_{\mrs,0}$ is independent of $z$. Multiplying \eqref{eq:cons_mass_s_O1_a} by $\Phi_{\mrs,0}$ and integrating it over the $\Omega_\mrs$, we have that
\begin{equation}
\int_{\Omega_\mrs} \nabla_z \cdot \left( \sigma_\mrs \nabla_z \Phi_{\mrs,0} \right) \Phi_{\mrs,0} \dd V_z = 0,
\end{equation}
and applying the divergence theorem to the left hand side we find
\begin{equation}
-\int_{\Omega_\mrs} \sigma_\mrs |\nabla_z \Phi_{\mrs,0}|^2 \dd V_z + \int_{\partial \Omega_\mrs} \Phi_{\mrs,0}  \sigma_\mrs \nabla_z \Phi_{\mrs,0} \cdot \vb*{n}_s \dd A_z = 0.
\end{equation}
The boundary integral over $\partial\Omega_\mrs$ vanishes because of the conditions \eqref{eq:cons_mass_s_O1_b} and \eqref{eq:cons_mass_s_O1_c}, and given that $\sigma_s$ is positive, $|\nabla_z \Phi_{\mrs,0}|^2 = 0$ over the entire domain $\Omega_\mrs$ which implies that $\Phi_{\mrs,0}$ is independent of $z$. This also means that $\tilde{\vb*{i}}_{\mrs,-1} = \vb*{0}$ and, therefore, the expansion of the current starts at $\order{1}$ as one would expect.

\subsubsection{$O(\delta)$ problem}
Now we consider the problem at $\order{\delta}$, which, using the results from the $\order{1}$ problem, is given by
\begin{subequations}\label{eq:cons_mass_s_Odelta}
	\begin{align}
	\nabla_z \cdot \tilde{\vb*{i}}_{\mrs,0} &= 0, & \text{ in } \Omega_\mrs,\\
	\tilde{\vb*{i}}_{\mrs,0} \cdot \vb*{n}_\mrs &= 0, & \text{ at } \partial\Omega_\mrs^{\mathrm{in}},\\
	& \text{periodic}, & \text{ at } \partial\Omega_\mrs^{\mathrm{out}},
	\end{align}
	with
	\begin{align}
	\tilde{\vb*{i}}_{\mrs,0} &= - \sigma_\mrs \left(\nabla_y \Phi_{\mrs,0} +  \nabla_z \Phi_{\mrs,1}\right).
	\end{align}    
\end{subequations}
Following the usual procedure for homogenisation (see \cite{Pavliotis2008} for details), we now write $\Phi_{\mrs,1} = \vb*{W}_\mrs \cdot \nabla_y \Phi_{\mrs,0}$, where $\vb*{W}_\mrs$ depends only on $z$. Substitution into \eqref{eq:cons_mass_s_Odelta} gives the cell problem
\begin{subequations}\label{eq:cell_pb_s}
	\begin{align}
	\nabla_z \cdot \left(\sigma_\mrs \left(\mathcal{I} + \nabla_z \vb*{W}_\mrs \right) \right) &= \vb*{0}, & \text{ in } \Omega_\mrs,\label{eq:cell_pb_s_a}\\
	\sigma_\mrs \left(\mathcal{I} + \nabla_z \vb*{W}_\mrs \right) \vb*{n}_\mrs &= \vb*{0}, & \text{ at } \partial\Omega_\mrs^{\mathrm{in}},\\
	& \text{periodic}, & \text{ at } \partial\Omega_\mrs^{\mathrm{out}},\\
	\int_{\Omega_\mrs} \vb*{W}_\mrs \dd V_z &= \vb*{0},
	\end{align}
\end{subequations}
which determines $\vb*{W}_\mrs$.

\subsubsection{$O(\delta^2)$ problem}
We finally address the $\order{\delta^2}$ problem which is given by
\begin{subequations}\label{eq:cons_mass_s_Odelta2}
	\begin{align}
	\nabla_y \cdot \tilde{\vb*{i}}_{\mrs,0} + \nabla_z \cdot \tilde{\vb*{i}}_{\mrs,1} &= 0, & \text{ in } \Omega_\mrs,\label{eq:cons_mass_s_Odelta2_a}\\
	\tilde{\vb*{i}}_{\mrs,1} \cdot \vb*{n}_\mrs &= G  \left(g_0 + C \dt\left(\Phi_{\mrs,0} - \Phi_{\mre,0} \right) \right), & \text{ at } \partial\Omega_\mrs^{\mathrm{in}},\label{eq:cons_mass_s_Odelta2_b}\\
	& \text{periodic}, & \text{ at } \partial\Omega_\mrs^{\mathrm{out}},\label{eq:cons_mass_s_Odelta2_c}
	\end{align}
	with
	\begin{align}
	\tilde{\vb*{i}}_{\mrs,1} &= - \sigma_\mrs \left(\nabla_y \Phi_{\mrs,1} +  \nabla_z \Phi_{\mrs,2}\right).
	\end{align}    
\end{subequations}
We now average \eqref{eq:cons_mass_s_Odelta2_a} over the domain $\Omega$ to determine the homogenised equations. Averaging the first term we find
\begin{equation}
\frac{1}{|\Omega|}\int_\Omega \nabla_y \cdot \tilde{\vb*{i}}_{\mrs,0} \dd V_z = \nabla_y \cdot \vb*{i}_{\mrs,0},
\end{equation}
where
\begin{equation}
\vb*{i}_{\mrs,0} = -\mathcal{S} \nabla_y \Phi_{\mrs,0}
\end{equation}
is the homogenised current, and
\begin{equation}
\mathcal{S} = \frac{1}{|\Omega|}\int_\Omega \sigma_\mrs \left(\mathcal{I} + \left(\nabla_z \vb*{W}_\mrs\right)^\mathrm{T} \right) \dd V_z
\end{equation}
is the electric conductivity tensor.

We can apply the divergence theorem to the second term of \eqref{eq:cons_mass_s_Odelta2_a} jointly with the conditions \eqref{eq:cons_mass_s_Odelta2_b} and \eqref{eq:cons_mass_s_Odelta2_c} to obtain
\begin{equation}
\frac{1}{|\Omega|}\int_\Omega \nabla_z \cdot \tilde{\vb*{i}}_{\mrs,1} \dd V_z = \frac{1}{|\Omega|}\int_{\partial\Omega_\mrs^\mathrm{in}} G  \left(g_0 + C \dt\left(\Phi_{\mrs,0} - \Phi_{\mre,0} \right) \right)  \dd A_z.
\end{equation}

Therefore, we conclude
\begin{subequations}\label{eq:homogenised_cons_mass_s}
	\begin{equation}
	\nabla_y \cdot \vb*{i}_{\mrs,0} = -J,
	\end{equation}
	where
	\begin{equation}
	J = \frac{1}{|\Omega|}\int_{\partial\Omega_\mrs^\mathrm{in}} G  \left(g_0 + C \dt\left(\Phi_{\mrs,0} - \Phi_{\mre,0} \right) \right)  \dd A_z.
	\end{equation}
\end{subequations}

\subsection{Conservation of mass and charge in the electrolyte}\label{sec:electrolyte}
We now focus on the equations in the electrolyte presented in \eqref{eq:electrolyte_micro}. Splitting the differential operators into the two different scales we find
\begin{subequations}
	\begin{align}
	\dt c_\mre + \nabla_y \cdot \tilde{\vb*{N}}_\mre + \frac{1}{\delta} \nabla_z \cdot \tilde{\vb*{N}}_\mre &= 0, & \text{ in } \Omega_\mre,\\
	\nabla_y \cdot \tilde{\vb*{i}}_\mre + \frac{1}{\delta} \nabla_z \cdot \tilde{\vb*{i}}_\mre &= 0, & \text{ in } \Omega_\mre,\\
	\tilde{\vb*{N}}_\mre \cdot \vb*{n}_\mre &= -\delta G  \left(g + C \dt\left(\Phi_\mrs - \Phi_\mre \right) \right), &  \text{ at } \partial\Omega_\mre^{\mathrm{in}},\\
	\tilde{\vb*{i}}_\mre \cdot \vb*{n}_\mre &= -\delta G  \left(g + C \dt\left(\Phi_\mrs - \Phi_\mre \right) \right), &  \text{ at } \partial\Omega_\mre^{\mathrm{in}},\\
	& \text{periodic}, & \text{ at } \partial\Omega_\mre^{\mathrm{out}},
	\end{align}
	with
	\begin{align}
	\tilde{\vb*{N}}_\mre &= -\left[D_\mathrm{L} \left( \nabla_y c_\mre + \frac{1}{\delta} \nabla_z c_\mre \right) + \lambda \mu_\mathrm{L} c_\mre \left(\nabla_y \Phi_\mre + \frac{1}{\delta} \nabla_z \Phi_\mre \right) \right],\\
	\tilde{\vb*{i}}_\mre &= -\left[(D_\mathrm{L} - D_\mathrm{A}) \left( \nabla_y c_\mre + \frac{1}{\delta} \nabla_z c_\mre \right) + \lambda (\mu_\mathrm{L} + \mu_\mathrm{A}) c_\mre \left(\nabla_y \Phi_\mre + \frac{1}{\delta} \nabla_z \Phi_\mre \right) \right].
	\end{align}
\end{subequations}

Now, we expand the following variables and fluxes as
\begin{subequations}
	\begin{align}
	c_\mre &= c_{\mre,0} + \delta c_{\mre,1} + \delta^2 c_{\mre,2} + \order{\delta^3},\\
	\Phi_\mre &= \Phi_{\mre,0} + \delta \Phi_{\mre,1} + \delta^2 \Phi_{\mre,2} + \order{\delta^3},\\
	\tilde{\vb*{N}}_\mre &= \tilde{\vb*{N}}_{\mre,0} + \delta \tilde{\vb*{N}}_{\mre,1} + \order{\delta^2},\\
	\tilde{\vb*{i}}_\mre &= \tilde{\vb*{i}}_{\mre,0} + \delta \tilde{\vb*{i}}_{\mre,1} + \order{\delta^2}.
	\end{align}
\end{subequations}
Even though the fluxes should have a term of $\order{\delta^{-1}}$, following a similar analysis as in Section \ref{sec:cons_mass_s_O1} we find that these components vanish. Then, the linearised problem yields the following equations. 

\subsubsection{$O(1)$ problem}
At leading order we have
\begin{subequations}
	\begin{align}
	-\left(D_\mathrm{L} \nabla_z c_{\mre,0} + \lambda \mu_\mathrm{L} c_{\mre,0} \nabla_z \Phi_{\mre,0} \right) &= \vb*{0},\\
	-\left((D_\mathrm{L} - D_\mathrm{A}) \nabla_z c_{\mre,0} + \lambda (\mu_\mathrm{L} + \mu_\mathrm{A}) c_{\mre,0} \nabla_z \Phi_{\mre,0} \right) &= \vb*{0}.
	\end{align}
\end{subequations}
If the diffusion coefficients and mobilities are independent of $c_{\mre,0}$, $\Phi_{\mre,0}$, and $z$, using a similar method as we did for $\Phi_{\mrs,0}$ we can show that $c_{\mre,0}$ and $\Phi_{\mre,0}$ do not depend on $z$. 

\subsubsection{$O(\delta)$ problem}
Using the results at $\order{1}$ to simplify the equations, at $\order{\delta}$ we have
\begin{subequations}\label{eq:electrolyte_Odelta}
	\begin{align}
	\nabla_z \cdot \tilde{\vb*{N}}_{\mre,0} &= 0, & \text{ in } \Omega_\mre,\\
	\nabla_z \cdot \tilde{\vb*{i}}_{\mre,0} &= 0, & \text{ in } \Omega_\mre,\\
	\tilde{\vb*{N}}_{\mre,0} \cdot \vb*{n}_\mre &= 0, &  \text{ at } \partial\Omega_\mre^{\mathrm{in}},\\
	\tilde{\vb*{i}}_{\mre,0} \cdot \vb*{n}_\mre &= 0, &  \text{ at } \partial\Omega_\mre^{\mathrm{in}},\\
	& \text{periodic}, & \text{ at } \partial\Omega_\mre^{\mathrm{out}},
	\end{align}
	with
	\begin{align}
	\tilde{\vb*{N}}_{\mre,0} &= -\left(D_\mathrm{L} \left(\nabla_y c_{\mre,0} + \nabla_z c_{\mre,1} \right) + \lambda \mu_\mathrm{L} c_{\mre,0} \left(\nabla_y \Phi_{\mre,0} + \nabla_z \Phi_{\mre,1} \right) \right),\\
	\tilde{\vb*{i}}_{\mre,0} &= -\left((D_\mathrm{L} - D_\mathrm{A}) \left(\nabla_y c_{\mre,0} + \nabla_z c_{\mre,1} \right) + \lambda (\mu_\mathrm{L} + \mu_\mathrm{A}) c_{\mre,0} \left(\nabla_y \Phi_{\mre,0} + \nabla_z \Phi_{\mre,1} \right) \right).
	\end{align}
\end{subequations}

We now write $c_{\mre,1} = \vb*{W}_\mre \cdot \nabla_y c_{\mre,0}$ and $\Phi_{\mre,1} = \vb*{V}_\mre \cdot \nabla_y \Phi_{\mre,0}$ so, using the fact that diffusivities and mobilities do not depend on $z$, \eqref{eq:electrolyte_Odelta} can be rearranged into the cell problems for each one. We find that both problems are identical, so we conclude that $\vb*{W}_\mre \equiv \vb*{V}_\mre$. In terms of notation we use $\vb*{W}_\mre$, which is determined by the cell problem
\begin{subequations}\label{eq:cell_pb_e1}
	\begin{align}
	\nabla_z \cdot \left(\mathcal{I} + \nabla_z \vb*{W}_\mre \right) &= \vb*{0}, & \text{ in } \Omega_\mre,\\
	\left(\mathcal{I} + \nabla_z \vb*{W}_\mre \right) \vb*{n}_\mre &= \vb*{0}, & \text{ at } \partial\Omega_\mre^{\mathrm{in}},\\
	& \text{periodic}, & \text{ at } \partial\Omega_\mre^{\mathrm{out}},\\
	\int_{\Omega_\mre} \vb*{W}_\mre \dd V_z &= \vb*{0}.
	\end{align}
\end{subequations}

\subsubsection{$O(\delta^2)$ problem}
Now we can consider the $\order{\delta^2}$ to determine the homogenised equations. The equations read
\begin{subequations}\label{eq:electrolyte_Odelta2}
	\begin{align}
	\dt c_{\mre,0} + \nabla_y \cdot \tilde{\vb*{N}}_{\mre,0} + \nabla_z \cdot \tilde{\vb*{N}}_{\mre,1} &= 0, & \text{ in } \Omega_\mre,\label{eq:electrolyte_Odelta2_a}\\
	\nabla_y \cdot \tilde{\vb*{i}}_{\mre,0} + \nabla_z \cdot \tilde{\vb*{i}}_{\mre,1} &= 0, & \text{ in } \Omega_\mre,\label{eq:electrolyte_Odelta2_b}\\
	\tilde{\vb*{N}}_{\mre,1} \cdot \vb*{n}_\mre &= - G  \left(g + C \dt\left(\Phi_\mrs - \Phi_\mre \right) \right), &  \text{ at } \partial\Omega_\mre^{\mathrm{in}},\label{eq:electrolyte_Odelta2_c}\\
	\tilde{\vb*{i}}_{\mre,1} \cdot \vb*{n}_\mre &= - G  \left(g + C \dt\left(\Phi_\mrs - \Phi_\mre \right) \right), &  \text{ at } \partial\Omega_\mre^{\mathrm{in}},\label{eq:electrolyte_Odelta2_d}\\
	& \text{periodic}, & \text{ at } \partial\Omega_\mre^{\mathrm{out}},\label{eq:electrolyte_Odelta2_e}
	\end{align}
	with
	\begin{multline}
	\tilde{\vb*{N}}_{\mre,1} = -\left(D_\mathrm{L} \left(\nabla_y c_{\mre,1} + \nabla_z c_{\mre,2} \right) \right. \\
	\left. + \lambda \mu_\mathrm{L} \left( c_{\mre,0} \left(\nabla_y \Phi_{\mre,1} + \nabla_z \Phi_{\mre,2} \right) + c_{\mre,1} \left(\nabla_y \Phi_{\mre,0} + \nabla_z \Phi_{\mre,1} \right) \right) \right),
	\end{multline}
	\begin{multline}
	\tilde{\vb*{i}}_{\mre,1} = -\left((D_\mathrm{L} - D_\mathrm{A}) \left(\nabla_y c_{\mre,0} + \nabla_z c_{\mre,1} \right) \right. \\
	\left. + \lambda (\mu_\mathrm{L} + \mu_\mathrm{A})  \left( c_{\mre,0} \left(\nabla_y \Phi_{\mre,1} + \nabla_z \Phi_{\mre,2} \right) + c_{\mre,1} \left(\nabla_y \Phi_{\mre,0} + \nabla_z \Phi_{\mre,1} \right) \right) \right).
	\end{multline}
\end{subequations}

We average \eqref{eq:electrolyte_Odelta2_a} and \eqref{eq:electrolyte_Odelta2_b} over $\Omega$ and use the divergence theorem with conditions \eqref{eq:electrolyte_Odelta2_c}-\eqref{eq:electrolyte_Odelta2_e} to obtain the homogenised equations
\begin{subequations}\label{eq:homogenised_electrolyte}
	\begin{align}
	\varphi_\mre \dt c_{\mre,0} + \nabla_y \cdot \vb*{N}_{\mre,0} &= J,\\
	\nabla_y \cdot \vb*{i}_{\mre,0} &= J,
	\end{align}
	where
	\begin{align}
	\vb*{N}_{\mre,0} &= -\left(D_{\mathrm{L}} \mathcal{B} \nabla_y c_{\mre,0} + \lambda \mu_L \mathcal{B} c_{\mre,0} \nabla_y \Phi_{\mre,0} \right),\\
	\vb*{i}_{\mre,0} &= -\left((D_{\mathrm{L}} - D_{\mathrm{A}}) \mathcal{B} \nabla_y c_{\mre,0} + \lambda (\mu_L + \mu_{\mathrm{A}}) \mathcal{B} c_{\mre,0} \nabla_y \Phi_{\mre,0} \right),\\
	\mathcal{B} &= \frac{1}{|\Omega|}\int_\Omega \left(\mathcal{I} + \left(\nabla_z \vb*{W}_\mre\right)^\mathrm{T} \right) \dd V_z.
	\end{align}
\end{subequations}

\subsection{Conservation of heat at the mesoscale}\label{sec:temperature_meso}
The temperature equation, splitting the microscale from larger scales, is given by
\begin{subequations}\label{eq:cons_energy_meso_multiscale}
	\begin{align}
	\rho c_p \dt T + \nabla_{y} \cdot \tilde{\vb*{K}} + \frac{1}{\delta} \nabla_z \cdot \tilde{\vb*{K}} &= - \lambda \tilde{\vb*{i}} \cdot \left( \nabla_y \Phi + \frac{1}{\delta} \nabla_z \Phi \right), & \quad \text{ in } \Omega,\\
	\left(\tilde{\vb*{K}}_\mrs - \tilde{\vb*{K}}_\mre \right) \cdot \vb*{n}_\mre &= \delta \lambda G (g \eta + g \Pi), & \quad \text{ in } \partial \Omega^{\mathrm{in}}_\mre,\\
	& \text{periodic}, & \quad \text{ at } \partial \Omega,
	\end{align}
	where
	\begin{equation}
	\tilde{\vb*{K}} = -k \left(\nabla_{y} T + \frac{1}{\delta} \nabla_z T \right).
	\end{equation}
\end{subequations}

Now we expand the following quantities as
\begin{subequations}
	\begin{align}
	T &= T_0 + \delta T_1 + \delta^2 T_2 + \order{\delta^3},\\
	\Phi &= \Phi_0 + \delta \Phi_1 + \delta^2 \Phi_2 + \order{\delta^3},\\
	\tilde{\vb*{K}} &= \tilde{\vb*{K}}_0 + \delta \tilde{\vb*{K}}_1 + \order{\delta^2},\\
	\tilde{\vb*{i}} &= \tilde{\vb*{i}}_0 + \delta \tilde{\vb*{i}}_1 + \order{\delta^2},
	\end{align}
\end{subequations}
where again it can be shown that the $\order{\delta^{-1}}$ contributions in the fluxes vanish.

\subsubsection{$O(1)$ problem}
At leading order the problem reads
\begin{equation}\label{eq:cons_energy_meso_O1}
	-k \nabla_z T_0 = \vb*{0}.
\end{equation}
Therefore, following a similar argument to the one for $\Phi_{\mrs,0}$ we conclude that $T_0$ is independent of $z$.

\subsubsection{$O(\delta)$ problem}
We now consider the $\order{\delta}$ problem, and using that neither $T_0$ nor $\Phi_0$ depend on $z$, we have
\begin{subequations}\label{eq:cons_energy_meso_Odelta}
	\begin{align}
	\nabla_z \cdot \tilde{\vb*{K}}_0 &= 0, & \quad \text{ in } \Omega,\\
	\left(\tilde{\vb*{K}}_{\mrs,0} - \tilde{\vb*{K}}_{\mre,0} \right) \cdot \vb*{n}_\mre &= 0, & \quad \text{ in } \partial \Omega^{\mathrm{in}}_\mre,\\
	& \text{periodic}, & \quad \text{ at } \partial \Omega,
	\end{align}
	where
	\begin{equation}
	\tilde{\vb*{K}}_0 = -k \left(\nabla_y T_0 + \nabla_z T_1 \right).
	\end{equation}
\end{subequations}
Defining $T_1 = \vb*{W}_T \cdot \nabla_y T_0$ we obtain the cell problem for the thermal problem
\begin{subequations}\label{eq:cell_pb_T}
	\begin{align}
	\nabla_z \cdot \left(k \left(\mathcal{I} + \nabla_z \vb*{W}_T \right) \right) &= \vb*{0}, & \quad \text{ in } \Omega,\label{eq:cell_pb_T_a}\\
	& \text{periodic}, & \quad \text{ at } \partial \Omega,\\
	\int_\Omega \vb*{W}_T \dd V_z &= \vb*{0},
	\end{align}
\end{subequations}
and notice that we do not need to account for the condition at the internal boundary as \eqref{eq:cell_pb_T_a} takes care of it.

\subsubsection{$O(\delta^2)$ problem}
We finally consider the $\order{\delta^2}$ problem
\begin{subequations}\label{eq:cons_energy_meso_Odelta2}
	\begin{align}
	\rho c_p \dt T_0 + \nabla_y \cdot \tilde{\vb*{K}}_0 + \nabla_z \cdot \tilde{\vb*{K}}_1 &= - \lambda \tilde{\vb*{i}}_0 \cdot \left( \nabla_y \Phi_0 + \nabla_z \Phi_1 \right), & \quad \text{ in } \Omega,\label{eq:cons_energy_meso_Odelta2_a}\\
	\left(\tilde{\vb*{K}}_{\mrs,1} - \tilde{\vb*{K}}_{\mre,1} \right) \cdot \vb*{n}_\mre &= \lambda G (g_0 \eta_0 + g_0 \Pi_0), & \quad \text{ in } \partial \Omega^{\mathrm{in}}_\mre,\\
	& \text{periodic}, & \quad \text{ at } \partial \Omega,
	\end{align}
	where
	\begin{equation}
	\tilde{\vb*{K}} = -k \left(\nabla_y T_1 + \nabla_z T_2 \right).
	\end{equation}
\end{subequations}
We can now average \eqref{eq:cons_energy_meso_Odelta2_a} over $\Omega$ to obtain the homogenised equation, doing the same type of manipulations as detailed in Section~\ref{sec:cons_charge_solid}, The integral can be split up into one integral over $\Omega_{s}$ and one over $\Omega_{e}$, so it can be written as
\begin{equation}
\begin{aligned}
\int_{\Omega} \vb*{i}_{0} \cdot (\nabla_{y}\Phi_{0}+\nabla_{z}\Phi_{1}) \dd V_z &= \int_{\Omega_\mrs} \vb*{i}_{\mrs,0} \cdot (\nabla_{y}\Phi_{\mrs,0} + \nabla_{z}\Phi_{\mrs,1}) \dd V_z \\
& \quad + \int_{\Omega_\mre} \vb*{i}_{\mre,0} \cdot(\nabla_{y}\Phi_{\mre,0} + \nabla_{z}\Phi_{\mre,1}) \dd V_z \\
&= \int_{\Omega_\mrs} \vb*{i}_{\mrs,0} \cdot \left( (\mathcal{I} + \nabla_{z} \vb*{W}_\mrs) \nabla_{y}\Phi_{\mrs,0} \right) \dd V_z \\
& \quad + \int_{\Omega_\mre}\vb*{i}_{\mre,0} \cdot \left( (\mathcal{I} + \nabla_{z} \vb*{W}_\mre) \nabla_{y} \Phi_{\mre,0} \right) \dd V_z.
\end{aligned}
\end{equation}
The homogenised equation reads
\begin{subequations}\label{eq:cons_energy_meso_Odelta2_sols}
	\begin{equation}
	\theta \dt T_0 = \nabla_y \cdot \left( \mathcal{K} \nabla_y T_0 \right) + Q_\mrs + Q_\mre + Q_{\mathrm{irr}} + Q_{\mathrm{rev}},
	\end{equation}
	where
	\begin{align}
	Q_{\mrs} &= - \lambda \vb*{i}_{\mrs,0} \cdot \left(\mathcal{Q}_\mrs \nabla_y \Phi_{\mrs,0} \right),\\
	Q_{\mre} &= - \lambda \vb*{i}_{\mre,0} \cdot \left(\mathcal{Q}_\mre \nabla_y \Phi_{\mre,0} \right),\\
	Q_{\mathrm{irr}} &= \frac{1}{|\Omega |} \int_{\partial \Omega_\mre^\mathrm{in}} \lambda G g_0 \eta_0 \dd A_z,\\
	Q_{\mathrm{rev}} &= \frac{1}{|\Omega |} \int_{\partial \Omega_\mre^\mathrm{in}} \lambda G g_0 \Pi_0 \dd A_z,
	\end{align}
	and
	\begin{align}
	\theta &= \frac{1}{|\Omega |} \int_\Omega \rho c_p \dd V_z,\\
	\mathcal{K} &= \frac{1}{|\Omega|}\int_\Omega k \left(\mathcal{I} + \left(\nabla_z \vb*{W}_T\right)^\mathrm{T} \right) \dd V_z,\\
	\mathcal{Q}_\mrs &= \frac{\left(\mathcal{S}^{-1}\right)^\mathrm{T}}{|\Omega |} \int_\Omega \sigma_\mrs \left(\mathcal{I} + \nabla_z \vb*{W}_\mrs \right) \left(\mathcal{I} + \left(\nabla_z \vb*{W}_\mrs \right)^\mathrm{T} \right) \dd V_z,\\
	\mathcal{Q}_\mre &= \frac{\left(\mathcal{B}^{-1}\right)^\mathrm{T}}{|\Omega |} \int_\Omega \left(\mathcal{I} + \nabla_z \vb*{W}_\mre \right) \left(\mathcal{I} + \left(\nabla_z \vb*{W}_\mre \right)^\mathrm{T} \right) \dd V_z.
	\end{align}
\end{subequations}

\subsection{Conservation of heat at the macroscale}
We finally homogenise the heat equation at the mesoscale (cell level) to obtain the heat equation at the macroscale (battery level). The mesoscale equation comes from assembling three instances of \eqref{eq:cons_energy_meso_Odelta2_sols} to account for the electrodes and the separator, or even consider a more complicated structure to account for double coated electrodes as done in \cite{Hennessy2019}. However, given that temperature and heat flux must be continuous across the interface between parts we can write it as a single problem with space varying parameters. Then, we can split the scales as in \eqref{eq:splitting_nabla_eps} and write the mesoscale equation for temperature as
\begin{subequations}\label{eq:cons_energy_macro_multiscale}
	\begin{align}
	\theta \dt T + \nabla_{x} \cdot \vb*{K} + \frac{1}{\epsilon} \nabla_{y} \cdot \vb*{K} &= Q, & \quad \text{ in } \Omega_\mathrm{cell},\\
	& \text{periodic}, & \quad \text{ at } \partial \Omega_\mathrm{cell},
	\end{align}
	where
	\begin{equation}
	\vb*{K} = -\mathcal{K} \left(\nabla_{x} T + \frac{1}{\epsilon} \nabla_y T \right),
	\end{equation}
	and $Q$ is the heat generation term defined as
	\begin{equation}
	Q = \begin{cases}
	Q_\mathrm{s} + Q_\mathrm{e} + Q_{\mathrm{irr}} + Q_{\mathrm{rev}}, & \text{ in } \Omega_\mathrm{p} \text{ and } \Omega_\mathrm{n}, \\
	Q_\mathrm{e}, & \text{ in } \Omega_\mathrm{sep}.
	\end{cases}
	\end{equation}
	Notice that the gradients appearing in $Q$ are at the mesoscale only, so when we expand $Q$ in powers of $\epsilon$ no term of $\order{\epsilon^{-1}}$ arises.
\end{subequations}

We now expand the variables as
\begin{subequations}
	\begin{align}
	T &= T_0 + \epsilon T_1 + \epsilon^2 T_2 + \order{\epsilon^3},\\
	\vb*{K} &= \vb*{K}_0 + \epsilon \vb*{K}_1 + \order{\epsilon^2},
	\end{align}
\end{subequations}
substitute them into \eqref{eq:cons_energy_macro_multiscale} and linearise. 

\subsubsection{$O(1)$ problem}
At leading order we have
\begin{equation}\label{eq:cons_energy_macro_O1}
	-\mathcal{K} \nabla_y T_0 = \vb*{K}_{0},
\end{equation}
therefore, by the same argument as in Section \ref{sec:temperature_meso}, we conclude that $T_0$ does not depend on $y$. 

\subsubsection{$O(\epsilon)$ problem}
Using the results found at leading order, the $\order{\epsilon}$ problem reads
\begin{subequations}\label{eq:cons_energy_macro_Oepsilon}
	\begin{align}
	\nabla_{y} \cdot \vb*{K}_0 &= 0, & \quad \text{ in } \Omega_\mathrm{cell},\\
	& \text{periodic}, & \quad \text{ at } \partial \Omega_\mathrm{cell},
	\end{align}
	where
	\begin{equation}
	\vb*{K}_0 = -\mathcal{K} \left(\nabla_{x} T_0 + \nabla_y T_1 \right).
	\end{equation}
\end{subequations}
To obtain the cell problem, we make the substitution $T_1 = \vb*{W}_\mathrm{cell} \cdot \nabla_x T_0$ finding
\begin{subequations}\label{eq:cell_pb_Tbatt}
	\begin{align}
	\nabla_y \cdot \left(\mathcal{K} \left(\mathcal{I} + \nabla_y \vb*{W}_\mathrm{cell} \right) \right) &= \vb*{0}, & \quad \text{ in } \Omega_\mathrm{cell},\\
	& \text{periodic}, & \quad \text{ at } \partial \Omega_\mathrm{cell},\\
	\int_{\Omega_\mathrm{cell}} \vb*{W}_\mathrm{cell} \dd V_y &= \vb*{0}.
	\end{align}
\end{subequations}

\subsubsection{$O(\epsilon^2)$ problem}
Finally, we consider the $\order{\epsilon^2}$ problem given by
\begin{subequations}\label{eq:cons_energy_macro_Oepsilon2}
	\begin{align}
	\theta \dt T_0 + \nabla_{x} \cdot \vb*{K}_0 + \nabla_{y} \cdot \vb*{K}_1 &= Q_0, & \quad \text{ in } \Omega_\mathrm{cell},\\
	& \text{periodic}, & \quad \text{ at } \partial \Omega_\mathrm{cell},
	\end{align}
	where
	\begin{equation}
	\vb*{K}_1 = -\mathcal{K} \left(\nabla_{x} T_1 + \nabla_y T_2 \right).
	\end{equation}
\end{subequations}
Averaging over $\Omega_\mathrm{cell}$ we obtain the homogenised equation
\begin{subequations}
	\begin{equation}
	\theta_\mathrm{batt} \dt T_0 = \nabla_x \cdot \left(\mathcal{K}_\mathrm{batt} \nabla_x T_0 \right) + Q_\mathrm{batt},
	\end{equation}
	where
	\begin{align}
	\theta_\mathrm{batt} &= \frac{1}{|\Omega_\mathrm{cell}|} \int_{\Omega_\mathrm{cell}} \theta \dd V_y,\\
	\mathcal{K}_\mathrm{batt} &= \frac{1}{|\Omega_\mathrm{cell}|} \int_{\Omega_\mathrm{cell}} \mathcal{K} \left(\mathcal{I} + \left(\nabla_y \vb*{W}_\mathrm{cell}\right)^\mathrm{T} \right) \dd V_y,\\
	Q_\mathrm{batt} &= \frac{1}{|\Omega_\mathrm{cell}|} \int_{\Omega_\mathrm{cell}} Q \dd V_y.
	\end{align}
\end{subequations}

\subsection{Discussion of the effective equations}
We observe a few key differences between the microscale model presented in Section \ref{sec:microscale} and the homogenised one derived in this section. The first one is that, even though the microscale equations assumed isotropic materials and therefore all the transport properties were defined as scalars, in the homogenised equations the same transport properties become tensors and thus allow for anisotropy. This anisotropy arises from the porous structure. In some cases, such as with the electronic conductivity $\mathcal{S}$, the anisotropy is caused by both geometric effects and the inhomogeneities in the material properties. However, in the electrolyte properties, the anisotropy arises purely from the geometry, which allows us to define the tensor $\mathcal{B}$ which accounts for the anisotropy of all electrolyte properties. Therefore, all the electrolyte transport tensors at the mesoscale must be multiples of each other.

Another difference is the appearance of source terms in the electrochemical equations. These source terms capture the chemical reaction that, at the microscale, occurs at the boundary between electrode and electrolyte. However, due to the homogenisation process, this boundary is lost at the mesoscale but the reactions are captured by the source term. The same happens with the reaction contribution to the heat source terms $Q_\mathrm{irr}$ and $Q_\mathrm{rev}$.

In the thermal equations, both at mesoscale and macroscale, we observe that what was the $\rho c_p$ term in the microscale equation became the averaged volumetric heat capacities $\theta$ and $\theta_\mathrm{batt}$, which can no longer be split into the averaged density and the averaged specific heat capacity. We also notice that the heat generation terms at the mesoscale have a similar structure to those at the microscale, but with the introduction of two tensors $\mathcal{Q}_\mrs$ and $\mathcal{Q}_\mre$ that account for microstructure effects in Joule heating.

The macroscale thermal equation holds for any periodic cell structure, however notice that normally we encounter layered materials. In that case, as it is well known from the literature \cite{Pavliotis2008}, we find that the tensor $\mathcal{K}_\mathrm{batt}$ is a diagonal tensor, and the diagonal values are the arithmetic average of the conductivities of each layer in the directions parallel to the layers, and the harmonic average of the conductivities in the direction perpendicular to the layers. Therefore, asymptotic homogenisation provides a rigorous proof to a result that is commonly used in the literature (e.g. \cite{Mei2019}).

For the sake of clarity, when performing the homogenisation we have not explicitly accounted for the temperature dependence on the parameters, even though the parameters are allowed to depend on temperature. Given that we have determined that, at leading order, temperature does not depend on $z$, the analysis for temperature dependent parameters follows exactly the same way with some extra parameters arising from the derivatives of the parameters with respect to temperature.

\subsection{Properties of the tensors}\label{sec:properties_tensors}
In the previous sections we have derived tensors that account for the material properties in the homogenised problem. These tensors have certain properties that show the type of behaviour to expect from the homogenised equations.

\subsubsection{Symmetry}
We start showing that the tensors are symmetric, which can be used to simplify some of the expressions. We show it for $\mathcal{S}$ because it has the most complicated definition in our model, and any other tensor that we have defined can be thought to be, from a mathematical point of view, a particular case of $\mathcal{S}$. The method used here follows closely the one used in \cite{Richardson2012}. For simplicity in the notation, we define $W_\mrs^{(i)}$ to be the $i$-th component of the vector $\vb*{W}_\mrs$ and $\vb*{e}_i$ is the basis vector in the $i$-th direction. Then, we define the integral
\begin{equation}
\int_{\Omega_\mrs} \nabla_z \cdot \left( \sigma_\mrs W_\mrs^{(i)} \left(\vb*{e}_j +\nabla_z W_\mrs^{(j)} \right) - \sigma_\mrs W_\mrs^{(j)} \left(\vb*{e}_i +\nabla_z W_\mrs^{(i)} \right) \right) \dd V_z = 0,
\end{equation}
which we can show to be zero using the divergence theorem and the boundary conditions in \eqref{eq:cell_pb_s}. Expanding the divergence in the integral, and using \eqref{eq:cell_pb_s_a} to eliminate some of the terms, we find
\begin{equation}
\int_{\Omega_\mrs} \sigma_\mrs \nabla_z W_\mrs^{(i)} \vb*{e}_j - \sigma_\mrs \nabla_z W_\mrs^{(j)} \vb*{e}_i \dd V_z = 0,
\end{equation}
and therefore we have 
\begin{equation}
\int_{\Omega_\mrs} \sigma_\mrs \pdv{W_\mrs^{(i)}}{z_j} \dd V_z = \int_{\Omega_\mrs} \sigma_\mrs \pdv{W_\mrs^{(j)}}{z_i} \dd V_z,
\end{equation}
from which we deduce that $\mathcal{S}$ is symmetric. Using similar procedures we can show that the other tensors we defined ($\mathcal{B}$, $\mathcal{K}$, $\mathcal{K}_\mathrm{batt}$, $\mathcal{M}_\mrs$ and $\mathcal{M}_\mre$) are symmetric as well.

\subsubsection{Reduction of the transport properties}
We now want to show that the presence of the microstructure reduces the transport properties. Mathematically, this is equivalent to showing that the diagonal elements of the tensor of a given transport phenomenon are smaller than the average of the microscale value over the cell for that same property. By considering the integral
\begin{equation}
\int_{\Omega_\mrs} \nabla_z \cdot \left(\sigma_\mrs \left(\vb*{e}_i + \nabla_z W_\mrs^{(i)} \right) \right) \dd V_z = 0,
\end{equation}
and manipulating it in a similar way it can be shown that the diagonal entries of the tensor satisfy
\begin{equation}
\mathcal{S}_{ii} < \frac{1}{|\Omega_\mrs |} \int_{\Omega_\mrs} \sigma_\mrs \dd V_z.
\end{equation}

\subsubsection{Analytical representation of the tensors}
Exploiting the symmetry of the tensors we can simplify the way we defined some of them. Thus, we present the simpler expressions here under the same section for convenience of the reader.

The electric conductivity tensor in the electrode is given by
\begin{equation}
\mathcal{S} = \frac{1}{|\Omega|}\int_\Omega \sigma_\mrs \left(\mathcal{I} + \left(\nabla_z \vb*{W}_\mrs\right)^\mathrm{T} \right) \dd V_z,
\end{equation}
where the cell variable $\vb*{W}_\mrs$ solves \eqref{eq:cell_pb_s}.

The geometry of the electrolyte is captured by the tensor $\mathcal{B}$ which can then be used to define the transport properties in the electrolyte. The tensor is defined by
\begin{equation}
\mathcal{B} = \frac{1}{|\Omega|}\int_\Omega \left(\mathcal{I} + \left(\nabla_z \vb*{W}_\mre\right)^\mathrm{T} \right) \dd V_z,
\end{equation}
where the cell variable $\vb*{W}_\mre$ solves \eqref{eq:cell_pb_e1}. Using $\mathcal{B}$ we can define the tensors for diffusivities and mobilities used in \eqref{eq:cons_charge_meso} as
\begin{align}
\mathcal{D}_\mathrm{L} &= D_\mathrm{L} \mathcal{B}, & \mathcal{D}_\mathrm{A} &= D_\mathrm{A} \mathcal{B}, & \mathcal{M}_\mathrm{L} &= \mu_\mathrm{L} \mathcal{B}, & \mathcal{M}_\mathrm{A} &= \mu_\mathrm{A} \mathcal{B}.
\end{align}

The thermal conductivity of the cell is given by
\begin{equation}
\mathcal{K} = \frac{1}{|\Omega|}\int_\Omega k \left(\mathcal{I} + \left(\nabla_z \vb*{W}_T\right)^\mathrm{T} \right) \dd V_z,
\end{equation}
where the cell variable $\vb*{W}_T$ solves \eqref{eq:cell_pb_T}, and the Joule heating tensors in the electrode and the electrolyte are defined as
\begin{align}
\mathcal{Q}_\mrs &= \frac{\mathcal{S}^{-1}}{|\Omega |} \int_\Omega \sigma_\mrs \Big(\mathcal{I} + \left(\nabla_z \vb*{W}_\mrs \right)\Big) \left(\mathcal{I} + \left(\nabla_z \vb*{W}_\mrs \right)^\mathrm{T} \right) \dd V_z,\\
\mathcal{Q}_\mre &= \frac{\mathcal{B}^{-1}}{|\Omega |} \int_\Omega \Big(\mathcal{I} + \left(\nabla_z \vb*{W}_\mre \right) \Big) \left(\mathcal{I} + \left(\nabla_z \vb*{W}_\mre \right)^\mathrm{T} \right) \dd V_z.    
\end{align}

Finally, the thermal conductivity of the battery is given by
\begin{equation}
\mathcal{K}_\mathrm{batt} = \frac{1}{|\Omega|_\mathrm{cell}}\int_{\Omega_\mathrm{cell}} \mathcal{K} \left(\mathcal{I} + \left(\nabla_y \vb*{W}_\mathrm{cell}\right)^\mathrm{T} \right) \dd V_y,
\end{equation}
where the cell variable $\vb*{W}_\mathrm{cell}$ solves \eqref{eq:cell_pb_Tbatt}.

\section{Comparison with the DFN model}\label{sec:comparison_DFN}
The model presented here can be regarded as a generalised version of the well-known DFN model \cite{Fuller1994}. In this section we compare both models and point out the main differences. The DFN model does not include the battery level nor the thermal model, so the comparison will be only at the micro and mesoscale and at constant temperature. Another difference betwen the model presented here and the DFN model is that the first considers a capacitance term in the exchange current between the electrode and the electrolyte, thus we can reduce to the DFN model by setting $C = 0$.

\subsection{Comparison at the microscale}
One of the key assumptions of the DFN model is that the electrode particles are spherical and that there is a single representative particle at each point of the mesoscale electrode (i.e. one particle in each homogenisation cell). With these assumptions, our model \eqref{eq:cons_mass_s_homog} reduces to
\begin{subequations}
\begin{align}
\dt c_\mrs &= \frac{1}{r^{2}}\pdv{}{r}\left(r^{2}D_\mrs \pdv{c_\mrs}{r}\right), & \text{ in } 0 < r < R,\\
- D_\mrs \pdv{c_\mrs}{r} &= G g, & \text{ at } r = R,\\
\pdv{c_\mrs}{r} &= 0, & \text{ at } r = 0,
\end{align}
\end{subequations}
which is the same as in the DFN model. Here, $r$ is the microscale space variable and takes the role of $z$ in our model.

\subsection{Comparison at the mesoscale}
Note that the parameters in the DFN model are all scalar, whereas in the model presented in this paper some parameters are tensors, which can take into account anisotropy. However, it can be reduced to the isotropic case (as in the DFN model) by taking the tensors to be multiples of the identity tensor, as then they can be replaced by a scalar.

Another main difference is the difference in dilute and concentrated theory which will necessarily lead to different expressions. However, the analysis presented here can be applied to the concentrated electrolyte equations (see \cite{Arunachalam2015,Ciucci2011} for details) to obtain analogous results.

\subsubsection{Conservation of charge in the electrode}
The conservation of charge for the solid in our model \eqref{eq:cons_charge_meso} reads
\begin{equation}
\nabla\cdot(\mathcal{S} \nabla \Phi_\mrs) = J.
\end{equation}
Due to the spherical symmetry of the particles assumed in the DFN model, we can write the exchange current as
\begin{equation}
    J = a G \left(g + C \dt (\Phi_\mrs - \Phi_\mre) \right),
\end{equation}
where $a$ is the surface area of the particles per unit of volume. To reduce to the DFN model, we need to neglect the capacitance of the double layer and the anistropy of the material.

Notice that, even though we reduced our model to the DFN model, in fact, if we take the microstructure from the latter (a single particle surrounded by electrolyte) we will find that the conductivity is zero as the particles are not in contact with each other. In the DFN model, this is circumvented by taking an effective value for the conductivity instead of calculating it from the microscale problem.

\subsubsection{Conservation of mass and charge in the electrolyte}
Finally, we consider the equations in the electrolyte. Here is where the main difference arises: the DFN model uses concentrated electrolyte theory while we use dilute electrolyte theory (see \cite{Newman2004} for details on both). Therefore, we cannot directly reduce the model presented here to the DFN model, but we can show that they have similar forms. This is useful as it points out that the concentrated electrolyte equations could be homogenised following a very similar method to the one presented here.

We start assuming isotropy, so we can take the coefficients to be scalars. We define $\mathcal{D}_{L}=D_{L} \mathcal{B}$, $\mathcal{D}_{A}=D_{A} \mathcal{B}$, $\mathcal{M}_{L}=\mu_{L} \mathcal{B}$, and $\mathcal{M}_{A}=\mu_{A} \mathcal{B}$, and due to isotropy we have $\mathcal{B} = B\mathcal{I} $ where $B$ is a scalar that captures the microstructure effect on the transport properties. For example, if we used the Bruggeman correlation, then $B = \varphi_\mre ^{1.5}$ (see \cite{Tjaden2016} for details). The transference number of the lithium ions is defined as 
\begin{equation}
    t^+ = \frac{\mu_\mathrm{L}}{\mu_\mathrm{L} + \mu_\mathrm{A}},
\end{equation}
and given that for dilute electrolyte theory the mobilities and diffusivities are proportional to each other, we can replace the mobilities in the definition above for diffusivities. Multiplying \eqref{c_cons} by $t^+$ and substracting it from \eqref{l_conc} we find
\begin{equation}
\varphi_{e}\p_{t}c_{e}-\nabla\cdot\left(2 D_\mathrm{L} B (1-t^+) \nabla c_{e}\right)=(1-t^+) a G g.
\end{equation}

This is the same form as the DFN model by setting the effective ion diffusion coefficient to 
\begin{equation}
D_\mathrm{e,eff} = 2 D_\mathrm{L} B (1-t^+).
\end{equation}
As explained earlier, $B$ captures the microstructure effects. The $2 (1-t^+)$ factor captures the migration effects in the effective diffusion, but for the concentrated electrolyte this term should be $(1 - 2t^+)$. This discrepancy has been deeply discussed and resolved in \cite{Bizeray2016}.

For the conservation of charge equation we can define
\begin{align}
    \kappa_\mathrm{D,eff} &= (D_\mathrm{L} - D_\mathrm{A}) c_\mre, & \kappa_\mathrm{eff} &= (\mu_\mathrm{L} + \mu_\mathrm{A}) c_\mre,
\end{align}
so \eqref{c_cons}, together with \eqref{eq:def_ie}, can be rearranged into 
\begin{equation}
\nabla \cdot \left( \kappa_\mathrm{D,eff} \nabla \log c_\mre + \lambda \kappa_\mathrm{eff} \nabla \Phi_\mre \right) = - a G g,
\end{equation}
which is the same form presented in \cite{Plett2015}. However, this presents the same inconsistencies between dilute and concentrated electrolyte theories discussed in \cite{Bizeray2016}. Therefore, in order to obtain the effective equations for a concentrated electrolyte one can use the same homogenisation method used in this article with the concentrated electrolyte equations.

\section{Discussion}\label{sec:discussion}
We derived an effective model for the thermal-electrochemical behaviour of porous electrode batteries (with a particular interest in lithium ion batteries) using the method of asymptotic homogenisation. We started from the governing equations at the microscale, which impose mass and charge conservation both in the electrode and the electrolyte, with an ion intercalation reaction at the electrode-electrolyte interface modelled by the Butler-Volmer equation. We assumed that transport of lithium in the electrode is governed by diffusion only and that Ohm's law holds for the charge. In the electrolyte, we used Nernst-Planck equations to describe the transport phenomena, assuming thus a dilute electrolyte. For the thermal model we assumed heat is only driven by diffusion, and it is generated in the bulk of the material due to Joule heating and at the electrode-electrolyte interface due to the chemical reaction (we included both reversible and irreversible effects).

We exploited the disparity of length scales of the problem and took the limit of infinitely small electrode particles compared to electrode thickness. In this limit, we can derive the homogenised problem at the cell level (mesoscale). A particularity of the analysis is that we took the diffusion coefficient of lithium in the electrode to be small, in order to retrieve the diffusion of lithium only at the microscale, as observed in lithium ion batteries \cite{Ecker2015ii,Ecker2015i}. After deriving the mesoscale model for an electrode, we assembled the cell model (the model at the separator is a particular case of electrode model) and homogenised it to obtain the battery model, exploiting the limit of an infinitely thin cell compared to the battery.

The homogenised model presented here is a generalised version of the DFN model \cite{Fuller1994}, so widely used in the modelling literature. The main differences between our model and the DFN model are that our model includes thermal and capacitance effects, and can account for an arbitrary microstructure. Even though we have not explicitly introduced the temperature dependence of the parameters, as it would clutter the notation, the analysis presented here easily extends for that case and the same homogenised equations are obtained but with temperature dependent tensors. With respect to the microstructure, in our model we can determine the effective mesoscale parameters directly from the microscopic properties, instead of using theoretical or empirical correlations, such as the Bruggeman correlation \cite{Tjaden2016}, which are a subject of debate within the battery modelling field. The model presented here also extends the literature on derivation of multiscale models for batteries using asymptotic homogenisation \cite{Arunachalam2017,Arunachalam2015,Ciucci2011,Richardson2012} because it includes thermal effects. In addition, it also captures the diffusion of lithium at the microscale.

The model presented here can be used to simulate electrodes with more realistic geometries, given that it can deal with microstructures more complex than the spherical particles used in the DFN model. For example, one could use SEM images of real electrodes to define the geometry of the cell problems and then solve them numerically. The method used here can also be used in a very similar way to homogenise a concentrated electrolyte model. Another area of future work is to include the current collectors into the model. Also, our model can be used as a stepping stone towards including more complex physical effects that occur in batteries which depend both on electrochemistry and temperature, such as degradation. Several degradation mechanisms that occur in batteries have been considered in the literature \cite{Reniers2019}, but in many cases the degradation models have been coupled to the DFN model in a rather ad hoc way. The framework presented here could be extended to derive degradation models from the microscale models in a systematic and physically consistent way.

\appendix

\section{Non-dimensionalisation of the model}\label{sec:nondim}
In this section we introduce the dimensional model and the scalings that lead to the dimensionless microscale model presented in Section~\ref{sec:microscale}. The details of the derivation of the microscale equations can be found in the handbooks of the field (see \cite{Newman2004,Plett2015}). In the electrode we have mass transport driven by diffusion and charge transport follows Ohm's law, with the reaction at the electrode-electrolyte interface modelled by Butler-Volmer equation. We model the electrolyte as a dilute electrolyte, and therefore we use Nernst-Planck equation to describe the transport of both positive and negative electrodes (see \cite{Newman2004} for details). We also assume that the fluid is static. We can rearrange the equations to obtain an equation for the lithium ion and one equation for the current in the electrolyte. Finally, we model the temperature at the microscale using the heat equation. The model holds over both electrode and electrolyte domains and it has a source term in the bulk (Joule heating) and a source term at the electrode-electrolyte interface (reversible and irreversible reaction heating). 

The variables of the problem are the concentration of lithium in the electrodes $c_\mrs$, the concentration of lithium ions in the electrolyte $c_\mre$, the potentials in the electrode and electrolyte $\Phi_\mrs$ and $\Phi_\mre$, respectively, and the temperature $T$. For convenience in the notation, we define the fluxes, which are derived from the variables. We have the molar fluxes $\vb*{N}_\mrs$ and $\vb*{N}_\mre$, the currents $\vb*{i}_\mrs$ and $\vb*{i}_\mre$, and the heat fluxes $\vb*{K}_\mrs$ and $\vb*{K}_\mre$, defined in the electrode and the electrolyte respectively. For convenience in the notation, in this appendix we do not use tildes for the fluxes at the microscale as those are the only ones that appear.

The dimensional equations in the electrode are
\begin{subequations}\label{eq:electrode_micro_dim}
	\begin{align}
	\dt c_\mrs + \nabla \cdot \vb*{N}_\mrs &= 0, & \text{ in } \Omega_\mrs,\\
	\nabla \cdot \vb*{i}_\mrs &= 0, & \text{ in } \Omega_\mrs,\\
	F \vb*{N}_\mrs \cdot \vb*{n}_\mrs &= g + C_\Gamma \dt\left(\Phi_\mrs - \Phi_\mre \right), & \text{ at } \partial\Omega_\mrs^{\mathrm{in}},\\
	\vb*{i}_\mrs \cdot \vb*{n}_\mrs &= g + C_\Gamma \dt\left(\Phi_\mrs - \Phi_\mre \right), & \text{ at } \partial\Omega_\mrs^{\mathrm{in}},\\
	& \text{periodic}, & \text{ at } \partial\Omega_\mrs^{\mathrm{out}},
	\end{align}
	with
	\begin{align}
	\vb*{N}_\mrs &= - D_\mrs \nabla c_\mrs,\\
	\vb*{i}_\mrs &= - \sigma_\mrs \nabla \Phi_\mrs,
	\end{align}
	and
	\begin{align}
	g &= F K c_\mrs^\beta \left(1 - \frac{c_\mrs}{c_\mrs^{\max}} \right)^{1-\beta} c_\mre^{1-\beta} \left[\exp\left( (1-\beta) \frac{F}{R T} \eta \right) - \exp\left( - \beta \frac{F}{R T} \eta \right) \right],\\
	\eta &= \Phi_\mrs - \Phi_\mre - U_\mathrm{ocp}(c_\mrs),
	\end{align}
\end{subequations}
where $F$ is the Faraday constant, $C_\Gamma$ is the double layer capacitance, $D_\mrs$ is the diffusion coefficient of lithium in the electrode, and $\sigma_\mrs$ is the electronic conductivity of the electrode. We also have the exchange current at the electrode-electrolyte interface $g$, in which $K$ is the reaction rate, $c_\mrs^{\max}$ is the maximum concentration in the electrode, $R$ is the gas constant, $\beta$ is the transfer coefficient of the reaction, and $U_\mathrm{ocp}$ is the open circuit potential, which depends on the electrode concentration $c_\mrs$ evaluated at the interface with the electrolyte.

The dimensional equations for the electrolyte read
\begin{subequations}\label{eq:electrolyte_micro_dim}
	\begin{align}
	\dt c_\mre + \nabla \cdot \vb*{N}_\mre &= 0, & \text{ in } \Omega_\mre,\\
	\nabla \cdot \vb*{i}_\mre &= 0, & \text{ in } \Omega_\mre,\\
	F \vb*{N}_\mre \cdot \vb*{n}_\mre &= - \left(g + C_\Gamma \dt\left(\Phi_\mrs - \Phi_\mre \right) \right), &  \text{ at } \partial\Omega_\mre^{\mathrm{in}},\\
	\vb*{i}_\mre \cdot \vb*{n}_\mre &= - \left(g + C_\Gamma \dt\left(\Phi_\mrs - \Phi_\mre \right) \right), &  \text{ at } \partial\Omega_\mre^{\mathrm{in}},\\
	& \text{periodic}, & \text{ at } \partial\Omega_\mre^{\mathrm{out}},
	\end{align}
	with
	\begin{align}
	\vb*{N}_\mre &= -\left(D_\mathrm{L} \nabla c_\mre + F \mu_\mathrm{L} c_\mre \nabla \Phi_\mre \right),\\
	\vb*{i}_\mre &= -\left((D_\mathrm{L} - D_\mathrm{A}) \nabla c_\mre + F (\mu_\mathrm{L} + \mu_\mathrm{A}) c_\mre \nabla \Phi_\mre \right),
	\end{align}
\end{subequations}
where $D_\mathrm{L}$ is the diffusivity of lithium ions, $D_\mathrm{A}$ is the diffusivity of negative ions, $\mu_\mathrm{L}$ is the mobility of lithium ions, and $\mu_\mathrm{A}$ is the mobility of negative ions.

The governing equations for the thermal model are
\begin{subequations}\label{eq:temp_micro_dim}
	\begin{align}
	\rho c_p \dt T + \nabla \cdot \vb*{K} &= - \vb*{i} \cdot \nabla \Phi, & \quad \text{ in } \Omega,\label{eq:temp_micro_dim_a}\\
	\left(\vb*{K}_\mrs - \vb*{K}_\mre \right) \cdot \vb*{n}_\mre &= g (\eta + \Pi), & \quad \text{ in } \partial \Omega^{\mathrm{in}}_\mre,\\
	& \text{periodic}, & \quad \text{ at } \partial \Omega,
	\end{align}
	where
	\begin{equation}
	\vb*{K} = -k \nabla T,
	\end{equation}
\end{subequations}
where $\rho$ is the density, $c_p$ is the specific heat capacity, $k$ is the thermal conductivity, and $\Pi$ is the Peltier term, accounting for reversible heat generation. Given that the heat equation is defined both in the electrode and the electrolyte, we dropped the subscripts of $\vb*{K}$, $\vb*{i}$ and $\Phi$ in \eqref{eq:temp_micro_dim_a} to simplify notation.

We now scale the variables and derived quantities as
\begin{equation}
\begin{aligned}
x &= N L \hat{x}, & y &= L \hat{y}, & z &= \ell \hat{z}, & t &= t_0 \hat{t}, \\
c_\mrs &= c_0 \hat{c}_\mrs, & \vb*{N}_\mrs &= \frac{c_0 L}{t_0} \hat{\vb*{N}}_\mrs, & c_\mre &= c_0 \hat{c}_\mre, & \vb*{N}_\mre &= \frac{c_0 L}{t_0} \hat{\vb*{N}}_\mre, \\
\Phi_\mrs &= \Phi_0 \hat{\Phi}_\mrs, & \vb*{i}_\mrs &= i_0 \hat{\vb*{i}}_\mrs, & \Phi_\mre &= \Phi_0 \hat{\Phi}_\mre, & \vb*{i}_\mre &= i_0 \hat{\vb*{i}}_\mre, \\
T &= T_0 + \Delta T \; \hat{T}, & \vb*{K} &= \rho_0 c_{p0} \frac{\Delta T L}{t_0} \hat{\vb*{K}}_\mrs, & g &= g_0 \hat{g}, & \eta &= \Phi_0 \hat{\eta}, 
\end{aligned}
\end{equation}
and the parameters of the model are scaled as
\begin{equation}
\begin{aligned}
D_\mrs &= \frac{\ell^2}{t_0} \hat{D}_\mrs, & \sigma_\mrs &= \frac{i_0 L}{\Phi_0} \hat{\sigma}_\mrs, & U_\mathrm{ocp} &= \Phi_0 \hat{U}_\mathrm{ocp}, & \Pi &= \Phi_0 \hat{\Pi},\\
D_\mathrm{L} &= \frac{L^2}{t_0} \hat{D}_\mathrm{L}, & D_\mathrm{A} &= \frac{L^2}{t_0} \hat{D}_\mathrm{A}, & \mu_\mathrm{L} &= \frac{1}{R T_0} \frac{L^2}{t_0} \hat{\mu}_\mathrm{L}, & \mu_\mathrm{A} &= \frac{1}{R T_0} \frac{L^2}{t_0} \hat{\mu}_\mathrm{A},\\
c_\mrs^{\max} &= c_0 \hat{c}^{\max}_\mrs, & \rho &= \rho_0 \hat{\rho}, & c_{p} &= c_{p0} \hat{c}_p, & k &= \frac{t_0}{\rho_0 c_{p0} L^2} \hat{k}, 
\end{aligned}
\end{equation}
with
\begin{align}
t_0 &= \frac{F c_0 L}{i_0}, & \Delta T &= \frac{i_0}{L} \frac{R T_0}{F} \frac{t_0}{\rho_0 c_{p0}}, & g_0 &= F K c_0.
\end{align}
Here $N$ is the number of cells that compose a battery, $L$ is the thickness of a cell, $\ell$ is the typical length scale of the microstructure, $c_0$ is characteristic concentration, $\Phi_0$ is the characteristic potential, $i_0$ the characteristic current, $T_0$ is the reference temperature, $\rho_0$ is the characteristic density, and $c_{p0}$ is the characteristic heat capacity. We scale the differential operator $\nabla$ with the length scale $L$.

With these scalings, the following dimensionless numbers arise
\begin{equation}
\begin{aligned}
G &= \frac{g_0 L}{i_0 \ell}, & C &= \frac{C_\Gamma \Phi_0}{g_0 t_0}, & \lambda &= \frac{F \Phi_0}{R T_0}, & \gamma &= \frac{\Delta T}{T_0}, & \delta &= \frac{\ell}{L}, & \epsilon &= \frac{1}{N}.
\end{aligned}
\end{equation}

Using these scalings we can nondimensionalise the model \eqref{eq:electrode_micro_dim}-\eqref{eq:temp_micro_dim}. Dropping hats in order to simplify the notation, we obtain the model \eqref{eq:electrode_micro}-\eqref{eq:temp_micro}.


%
%

\bibliographystyle{spphys}       
\bibliography{references}   

\end{document}